\documentclass[conference]{IEEEtran}
\usepackage{cite}
\usepackage{amsmath,amssymb,amsfonts}
\usepackage{algorithm}
\usepackage{graphicx}
\usepackage{textcomp}
\usepackage{xcolor}
\usepackage{xspace}
\usepackage{threeparttable, booktabs, tabularx}
\usepackage[hyphens]{url}
\usepackage{stmaryrd}
\usepackage{hyperref}
\hypersetup{
    colorlinks=true,
    linkcolor=magenta,
    filecolor=magenta,
    citecolor=DarkRed,
    urlcolor=cyan,
}
\usepackage[hyperpageref]{backref}
\usepackage{cleveref}
\usepackage{algpseudocode}
\usepackage{tablefootnote}

\newcommand{\ignore}[1]{}

\newcommand{\specialcell}[2][c]{%
\begin{tabular}[#1]{@{}c@{}}#2\end{tabular}}

\def\BibTeX{{\rm B\kern-.05em{\sc i\kern-.025em b}\kern-.08em
    T\kern-.1667em\lower.7ex\hbox{E}\kern-.125emX}}

\definecolor{DarkRed}{RGB}{150,0,0}
\newcommand{\vinod}[1]{}
\newcommand{\leo}[1]{}


\newcommand{\C}{\mathbb{C}}

\newcommand{\Z}{\mathbb{Z}}

\newcommand{\PtAdd}{\mathsf{PtAdd}}
\newcommand{\Add}{\mathsf{Add}}
\newcommand{\PtMult}{\mathsf{PtMult}}
\newcommand{\Automorph}{\mathsf{Automorph}}
\newcommand{\ModUp}{\mathsf{ModUp}}
\newcommand{\PModUp}{\mathsf{PModUp}}
\newcommand{\ModDown}{\mathsf{ModDown}}

\newcommand{\Mult}{\mathsf{Mult}}
\newcommand{\NewMult}{\mathsf{NewMult}}
\newcommand{\KeySwitch}{\mathsf{KeySwitch}}
\newcommand{\Rotate}{\mathsf{Rotate}}
\newcommand{\HRotate}{\mathsf{HRotate}}
\newcommand{\Conjugate}{\mathsf{Conjugate}}
\newcommand{\Bootstrap}{\mathsf{Bootstrap}}
\newcommand{\Rescale}{\mathsf{Rescale}}
\newcommand{\KSKInProd}{\mathsf{KSKInnerProd}}

\newcommand{\NewLimb}{\mathsf{NewLimb}}

\newcommand{\InnerProduct}{\mathsf{InnerProduct}}

\newcommand{\calB}{\mathcal{B}}

\newcommand{\PtMatVecMult}{\mathsf{PtMatVecMult}}

\newcommand{\PolyEval}{\mathsf{PolyEval}}

\newcommand{\NTT}{\mathsf{NTT}}
\newcommand{\iNTT}{\mathsf{iNTT}}

\newcommand{\ksk}{\mathsf{ksk}}
\newcommand{\rot}{\mathsf{rot}}

\newcommand{\CoeffToSlot}{\mathsf{CoeffToSlot}}
\newcommand{\SlotToCoeff}{\mathsf{SlotToCoeff}}

\newcommand{\sine}{\mathsf{sine}}
\newcommand{\fftIter}{\mathsf{fftIter}}

\newcommand{\dnum}{\mathsf{dnum}}

\newcommand{\Decomp}{\mathsf{Decomp}}
\renewcommand{\vec}[1]{\overrightarrow{#1}}

\newcommand{\throughput}{\mathsf{throughput}}
\newcommand{\brt}{\mathsf{brt}}
\newcommand{\bp}{\mathsf{bp}}

\newcommand{\dbrack}[1]{\left\llbracket #1 \right\rrbracket}

\renewcommand{\a}{\mathbf{a}}
\renewcommand{\b}{\mathbf{b}}
\renewcommand{\c}{\mathbf{c}}

\renewcommand{\k}{\mathbf{k}}
\newcommand{\M}{\mathbf{M}}
\newcommand{\m}{\mathbf{m}}
\newcommand{\s}{\mathbf{s}}
\renewcommand{\t}{\mathbf{t}}
\renewcommand{\u}{\mathbf{u}}
\renewcommand{\v}{\mathbf{v}}
\newcommand{\w}{\mathbf{w}}
\newcommand{\x}{\mathbf{x}}
\newcommand{\y}{\mathbf{y}}
\newcommand{\z}{\mathbf{z}}

\title{Does Fully Homomorphic Encryption\\ Need Compute Acceleration?}

\author{
	Leo de Castro$^{1,*}$
    Rashmi Agrawal$^{2,3,*,\$}$
	Rabia Yazicigil$^2$
	Anantha Chandrakasan$^1$\\
	Vinod Vaikuntanathan$^1$
	Chiraag Juvekar$^3$
	Ajay Joshi$^2$\\
	\small $^1$MIT, Cambridge, MA, USA;
	\small $^2$Boston University, Boston MA, USA;
	\small $^3$Analog Devices, Boston, MA USA\\
	\small \{ldec, anantha, vinodv\}@mit.edu, \{rashmi23, rty, joshi\}@bu.edu, chiraag.juvekar@analog.com\\
	\small$^*$Equal Contribution $^\$$Work done during internship at Analog Devices
	
	}

\begin{document}
\maketitle
\thispagestyle{plain}
\pagestyle{plain}


\begin{abstract}
The emergence of cloud-computing has raised important privacy questions about the data that users share with remote servers. While data in transit is protected using standard techniques like Transport Layer Security (TLS), most cloud providers have unrestricted plaintext access to user data at the endpoint. Fully Homomorphic Encryption (FHE) offers one solution to this problem by allowing for arbitrarily complex computations on encrypted data without ever needing to decrypt it. Unfortunately, all known implementations of FHE require the addition of \textit{noise} during encryption which grows during computation. As a result, sustaining deep computations requires a periodic noise reduction step known as {\em bootstrapping}. The cost of the bootstrapping operation is one of the primary barriers to the wide-spread adoption of FHE.

In this paper, we present an in-depth architectural analysis of the bootstrapping step in FHE.
First, we observe that secure implementations of bootstrapping exhibit a low arithmetic intensity ($<1$ Op/byte), require large caches ($>100$MB) and as such, are heavily bound by the main memory bandwidth. Consequently, we demonstrate that existing workloads observe marginal performance gains from the design of bespoke high-throughput arithmetic units tailored to FHE. 
Secondly, we propose several cache-friendly algorithmic optimizations that improve the throughput in FHE bootstrapping by enabling up to $3.2\times$ higher arithmetic intensity and $4.6\times$ lower memory bandwidth. Our optimizations apply to a wide range of structurally similar computations such as private evaluation and training of machine learning models. Finally, we incorporate these optimizations into an architectural tool which, given a cache size, memory subsystem, the number of functional units and a desired security level, selects optimal cryptosystem parameters to maximize the bootstrapping throughput.

Our optimized bootstrapping implementation represents a best-case scenario for compute acceleration of FHE. We show that despite these optimizations, bootstrapping (as well as other applications with similar computational structure) continue to remain bottlenecked by main memory bandwidth. We thus conclude that secure FHE implementations need to look beyond accelerated compute for further performance improvements and to that end, we propose new research directions to address the underlying memory bottleneck. In summary, our answer to the titular question is: {\em yes, but only after addressing the memory bottleneck}! 
\end{abstract}
\section{Introduction}
\label{section:Intro}
\begin{figure}[t]
  \begin{center}
    \includegraphics[width=1.00\columnwidth]{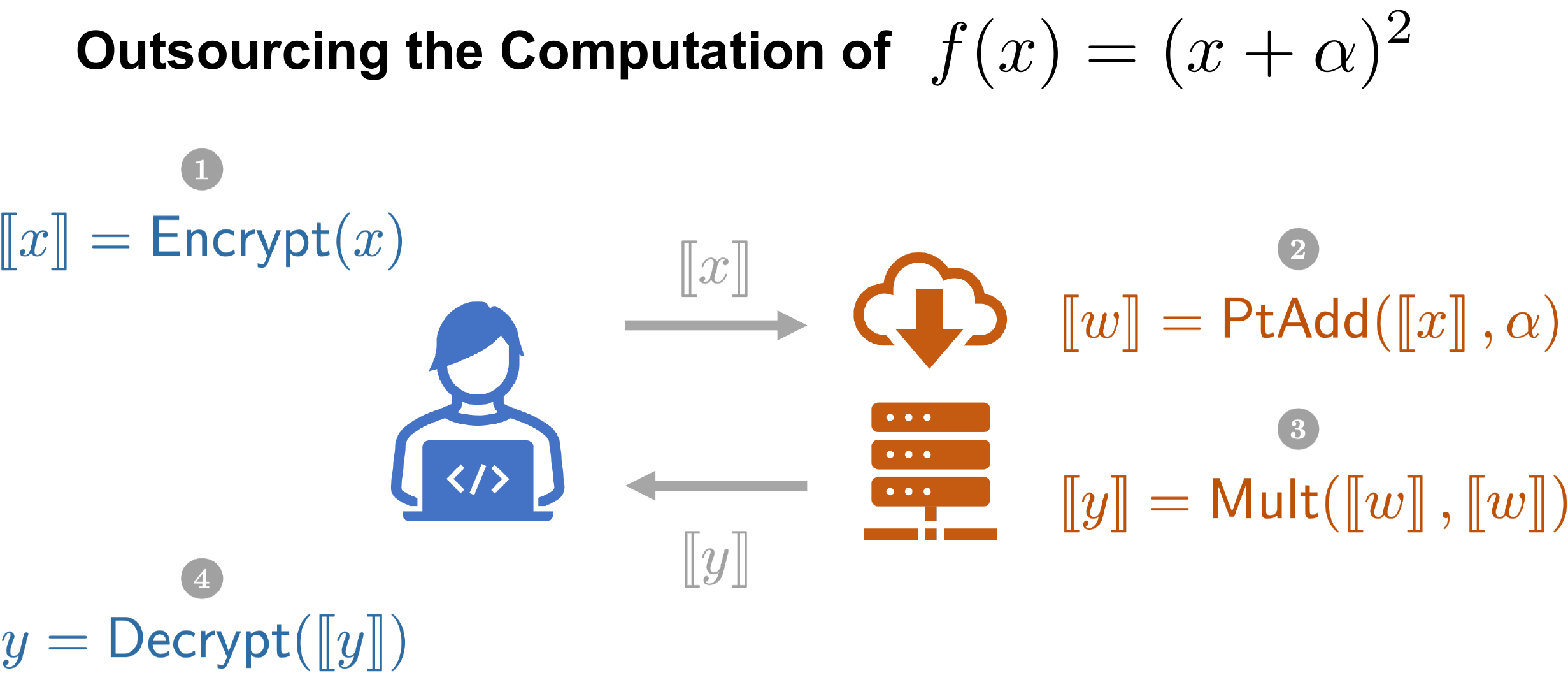} 
  \end{center}
  \vspace{-0.15in}
  \caption{Third-party cloud platform with outsourced FHE-based computing.}
  \vspace{-0.10in}
  \label{fig:cloud}
\end{figure}

The rapid development of cloud-based systems has enabled reliable and affordable access to shared computing resources at scale. 
However, this shared access raises substantial privacy and security challenges.
Therefore, new techniques are required to guarantee the confidentiality of sensitive user data when it is sent to the cloud for processing.
Fully Homomorphic Encryption (FHE)~\cite{RAD,Gentry09} enables cloud operators to perform complex computations on encrypted user data without ever needing to decrypt it.
The result of such FHE-based computation is in an encrypted form and can only be decrypted by the data owner. 
An illustrative use case of how a data owner can outsource computation on private data to an untrusted third-party cloud platform is shown in Figure~\ref{fig:cloud}.

While FHE-based privacy-preserving computing is promising, performing large encrypted computations with FHE still remains several orders of magnitude slower than operating on unencrypted data, which makes broad adoption impractical. 
This slowdown is an inherent feature of all existing lattice-based FHE schemes. 
All of these schemes produce ciphertexts containing a noise term, which is necessary for security. 
Each subsequent homomorphic operation performed on the ciphertext increases its noise, until it grows beyond a critical level after which recovery of the computation output is impossible. 
Sustained FHE computation thus requires a periodic de-noising procedure, called {\em bootstrapping}, to keep the noise below a correctness threshold. Unfortunately, this bootstrapping step is expensive in terms of both compute and memory requirements and is often $>100\times$ more expensive than primitive operations like addition and multiplication on encrypted data. 

Real-world applications commonly attempt to amortize this bootstrapping cost across multiple homomorphic operations. 
Even when considering these application-specific optimizations, bootstrapping consumes more than $50\%$ of the total compute and memory budget for end-to-end operations like machine learning training~\cite{GPUBoot21}. 
To make FHE-based computing practical, we need to consider a multi-layer approach to accelerate both the bootstrapping step as well as its primitive building blocks using a combination of algorithmic and hardware techniques.

In this work, we first perform a thorough compute and memory analysis of both simple and complex FHE primitives including the bootstrapping step, with an intent to determine the limits and potential opportunities for accelerating FHE. 
Our analysis reveals that all FHE operations exhibit low arithmetic intensity ($<1$ Op/byte) and require working-set sizes of hundreds of MB for practical and secure parameters. 
In fact, we observe that most existing performance optimization techniques for FHE often {\em increase} memory bandwidth requirements. 
These include both linear and non-linear operation optimizations proposed by Han and Ki~\cite{HK19}, Han, Hhan and Cheon~\cite{CHH18}, and Bossuat, Mouchet, Troncoso-Pastoriza and Hubaux~\cite{BMTH20}.
Recent bootstrapping implementation on GPUs by Jung, Kim, Ahn, Cheon and Lee~\cite{GPUBoot21} is the first work to perform memory-centric optimizations for linear operations in bootstrapping. 
Even after these optimizations, their implementation continues to be bounded by main memory bandwidth and exhibits an arithmetic intensity of $<1$ Op/byte. 
On the other side of the design spectrum, recent work by Samardzic et al.~\cite{F1Paper21} presents an architecture for a high-throughput hardware accelerator for FHE. 
This work primarily focuses on smaller parameter sets where full ciphertexts fit in on-chip cache memory allowing them to bypass the memory bandwidth limitation. 
However, many natural applications such as SIMD boostrapping, deep-neural network inference (with complex activation functions) and machine learning require larger parameter sets that are not addressed in \cite{F1Paper21}. 

In this work, we focus on presenting our three key contributions, i.e., application benchmarking, new techniques to improve memory performance, and evaluation of these techniques on end-to-end applications. More specifically:
\begin{itemize}
    \item We present detailed benchmarking of the compute and memory requirements of various FHE computations ranging from primitive operations to end-to-end applications such as machine-learning training. We show that all these benchmarks exhibit low arithmetic intensity and require large working-sets in on-chip memory. We observe that these working-sets do not fit in the last level caches of today's reticle-limited chips leading to bootstrapping and other applications being bottlenecked by memory accesses.
    \item We next present techniques to improve main memory bandwidth utilization by effectively managing the moderate last-level cache provided by currently available commercial hardware. For cache-pressured hardware ($<20$ MB LLC) we propose a domain-specific physical address mapping to enhance DRAM utilization. We then present hardware-independent algorithmic optimizations that reduce memory and compute requirements of FHE operations.
    \item We finally propose an optimized, memory-aware cryptosystem parameter set that maximizes the throughput in FHE bootstrapping and logistic regression training by enabling up to $3.2\times$ higher arithmetic intensity and $4.6\times$ lower memory bandwidth.
\end{itemize}

The techniques that we propose often compose with prior art and can be used as drop-ins to provide performance improvements in existing implementations without the need for new hardware.
Our proposed bootstrapping parameter set represents an upper limit on the performance of FHE operations that can be attained through pure compute acceleration when paired with existing state-of-art memory subsystems. 
Even with this optimal parameter set, we observe that the bootstrapping step is still primarily memory bound. Thus:

\begin{quote}
\textit{Our key conceptual take-away is that to accelerate FHE, we need novel techniques to address the underlying memory bandwidth issues. Compute acceleration alone is unlikely to make a dent.}
\end{quote}

Towards the goal of addressing memory bandwidth issues, we propose novel near-term algorithmic and architectural research directions.
\section{Fully Homomorphic Encryption: The API} 
\label{section:FHEAPI}
To set the stage, in this section we present the operations implemented by the Cheon-Kim-Kim-Song (CKKS)~\cite{CKKS17} FHE scheme. 
We organize these operations in the form of an API that can be used by any application developer to design privacy-preserving applications. 
Specifying the CKKS scheme requires several parameters, and we summarize our notation for these parameters in \Cref{tab:parameters}. 
Though we focus on the CKKS scheme, the API is generic and can be used for the BGV~\cite{BGV12} and B/FV~\cite{Brak12,FV12}  schemes as well\footnote{An exception is the $\Conjugate$ function, which the BGV and B/FV schemes do not support, since they do not encrypt complex numbers.}.

\begin{table}[t]
    \centering
    \caption{CKKS FHE Parameters and their description.}
    \label{tab:parameters}
    \begin{tabular}{p{0.15\columnwidth} p{0.7\columnwidth}}
    \toprule
    \textbf{Parameter} & \textbf{Description}\\
    \midrule
    $N$ & Number of coefficients in a polynomial in the ciphertext ring.\\
    $n$ & $N/2$, number of plaintext elements in a single ciphertext.\\
    $Q$ & Full modulus of a ciphertext coefficient.\\
    $q$ & Machine word sized prime modulus and a limb of $Q$.\\
    $\Delta$ & Scaling factor of a CKKS plaintext.  \\
    $P$ & Product of the additional limbs added for the raised modulus.\\
    $L$ & Maximum number of limbs in a ciphertext.\\
    $\ell$ & Current number of limbs in a ciphertext.\\
    $\dnum$ & Number of digits in the switching key.\\
    $\alpha$ & $\lceil (L + 1)/\dnum \rceil$. Number of limbs that comprise a single digit in the key-switching decomposition. This value is fixed throughout the computation. \\
    $\beta$ & $\lceil (\ell + 1)/\alpha \rceil$. An $\ell$-limb polynomial is split into this number of digits during base decomposition.\\
    \bottomrule
    \end{tabular}
\end{table}

\begin{table*}
    \centering
\begin{threeparttable}[t]
    \caption{CKKS Fully Homomorphic Encryption API.}
    \label{tab:low-level-api-overview}
    \begin{tabular}{ l l l p{\columnwidth}}
    \toprule
    \textbf{Operation Name} & \textbf{Output} & \textbf{Implementation} & \textbf{Description} \\
    \midrule
    $\PtAdd(\dbrack{\x}, \y)$ & $\dbrack{\x + \y}$ & $\dbrack{\x} + \y$ & Adds a plaintext vector to an encrypted vector. \\
    $\Add(\dbrack{\x}, \dbrack{\y})$ & $\dbrack{\x + \y}$ & $\dbrack{\x} + \dbrack{\y}$ & Adds two encrypted vectors.\\
    $\PtMult(\dbrack{\x}, \y)$ & $\dbrack{\x \cdot \y}$  & \Cref{algo:PtMult} & Multiplies a plaintext vector and an encrypted vector. \\
    $\Mult(\dbrack{\x}, \dbrack{\y})$ & $\dbrack{\x \cdot \y}$ & \Cref{algo:Mult} & Multiplies two encrypted vectors. \\ 
    $\Rotate(\dbrack{\x}, k)$ & $\dbrack{\phi_k(\x)}$ & \Cref{algo:Rotate} & Rotates a vector by $k$ positions; see \Cref{subsection:LowLevelAPIDef} for an illustration. \\
    $\Conjugate(\dbrack{\x})$ & $\dbrack{\overline{\x}}$ & \Cref{algo:Rotate}\tnote{$\dag$} & Outputs an encryption of the complex conjugate of the encrypted input vector. \\
    \bottomrule
    \end{tabular}
    \begin{tablenotes}
    \item[$\dag$] Through a clever encoding~\cite{CKKS17}, the $\Conjugate$ operation implementation is identical to the $\Rotate$ implementation.
    \end{tablenotes}
\end{threeparttable}
\end{table*}

\subsection{Homomorphic Encryption API} 
\label{subsection:LowLevelAPIDef}

The basic plaintext data-type in CKKS is a vector of length $n$ where each entry is chosen from $\C$, the field of complex numbers. 
All arithmetic operations on plaintexts are component-wise; the entries of the vector $\x + \y$ (resp. $\x \cdot \y$) are the component-wise sums (resp. products) of the entries of $\x$ with the corresponding entries of $\y$.
We denote the encryption of a length-$n$ vector $\x$ by $\dbrack{\x}$.

\Cref{tab:low-level-api-overview} gives a complete description of the API with the exception of the rotation operation, which we describe here. The $\Rotate$ operation takes in an encryption of a vector $\x$ of length $n$ and an integer $0 \leq k < n$, and outputs an encryption of a rotation of the vector $\x$ by $k$ positions. 
    As an example, when $k = 1$, the rotation $\phi_1(\x)$ is defined as follows. 
    \begin{align*}
        \x &= \begin{pmatrix}
        x_0 & x_1 & \ldots & x_{n-2} & x_{n-1}
        \end{pmatrix}\\
        \phi_1(\x) &=\begin{pmatrix}
        x_{n-1} & x_{0} & \ldots & x_{n-3} & x_{n-2}
        \end{pmatrix}
    \end{align*}
The $\Rotate$ operation is necessary for computations that operate on data residing in different slots of the encrypted vectors. 

\subsection{Modular Arithmetic and the Residue Number System}
\label{ssec:mod_arith}

\paragraph*{Scalar Modular Arithmetic}
Nearly all FHE operations reduce to scalar modular additions and scalar modular multiplications. 
Current CPU/GPU architectures do not implement modular arithmetic directly but emulate it via multiple arithmetic instructions, which significantly increases the amount of compute required for these operations. Therefore, optimizing modular arithmetic is critical to optimizing FHE computation.

To perform modular addition over operands that are already reduced, we use the standard approach of conditional subtraction if the addition overflows the modulus. For generic modular multiplications, we use the Barrett reduction technique~\cite{Barrett}. 
When computing the sum of many scalars, we avoid performing a modular reduction until the end of the summation, as long as the unreduced sum fits in a machine word. 
As an optimization, we use Shoup's technique~\cite{shoup2001ntl} 
for constant multiplication. 
That is, when computing $x\cdot y \pmod p$ where $x$ and $p$ are known in advance, we can precompute a value $x_s$ such that $\mathsf{ModMulShoup}(x, y, x_s, p) = x\cdot y \pmod p$ is much faster than directly computing $x\cdot y \pmod p$. 

\paragraph*{Residue Number System (RNS)}
Often the scalars in homomorphic encryption schemes are very large, on the order of thousands of bits.
To compute on such large numbers, we use the residue number system (also called the Chinese remainder representation) where we represent numbers modulo $Q = \prod_{i=1}^\ell q_i$, where each $q_i$ is a prime number that fits in a standard machine word (less than $64$ bits), as $\ell$ numbers modulo each of the $q_i$. 
We call the set $\calB := \{q_1, \ldots, q_\ell\}$ an \emph{RNS basis}. 
We refer to each $q_i$ as a \emph{limb} of $Q$.

This allows us to operate over values in $\Z_Q$ without any native support for multi-precision arithmetic. 
Instead, we can represent $x \in \Z_Q$ as a length-$\ell$ vector of scalars $[x]_\calB = (x_1, x_2, \ldots, x_\ell)$, where $x_i \equiv x \pmod{q_i}$. 
We refer to each $x_i$ as a \emph{limb} of $x$.
To add two values $x, y \in \Z_Q$, we have $x_i + y_i  \equiv x + y \pmod{q_i}$. Similarly, we have $x_i \cdot y_i  \equiv x \cdot y \pmod{q_i}$. 
This allows us to compute addition and multiplication over $\Z_Q$ while only operating over standard machine words. 
The size of this representation of an element of $\Z_Q$ is $\ell$ machine words.

\subsection{CKKS Ciphertext Structure} 
\label{subsec:CKKSStructure}
In this section, we give the general structure of a ciphertext in the CKKS~\cite{CKKS17} homomorphic encryption scheme. 
A ciphertext is a pair of polynomials each of degree $N-1$. 
The coefficients of these ciphertexts are elements of $\Z_Q$, where $Q$ has $\ell$ limbs. 
Thus, in total, the size of a ciphertext is $2N\ell$ machine words.

In CKKS, we are able to encrypt non-integer values, including complex numbers.
The ciphertexts are ``packed," which means they encrypt vectors in $\C^n$, where $n = N/2$, in a single ciphertext.
For $\m \in \C^n$, we denote its encryption as $\dbrack{\m} = (\a_\m, \b_\m)$ where $\a_\m$ and $\b_\m$ are  the two polynomials that comprise the ciphertext.
We omit the subscript $\m$ when there is no cause for confusion.

An example of ciphertext parameters that achieve a $128$-bit security level is $N = 2^{17}$ and $\ell = 35$.
With an $8$-byte machine word, this gives a total ciphertext size of $\sim 73.4$~MB. 
Note that in today's reticle-limited systems, the largest last-level cache size is $40$~MB~\cite{nvidiaA100}.
Consequently, we won't be able to fit even a single ciphertext in the last-level cache, which indicates the need for multiple expensive DRAM accesses when operating on ciphertexts.

\paragraph*{Polynomial Representation} In order to enable fast polynomial multiplication, we will have all polynomials represented by default as a series of $N$ evaluations at fixed roots of unity. 
This allows polynomial multiplication to occur in $O(N)$ time. 
We refer to this representation as the \emph{evaluation representation}. 
Certain subroutines, defined in \cref{subsec:CKKSSubRout}, operate over 
the polynomial's \emph{coefficient representation}, which is simply a vector of its coefficients. 
Addition of two polynomials and multiplication of a polynomial by a scalar are $O(N)$ in both the coefficient and the evaluation representation. 
Moving between representations requires a number-theoretic transform 
(NTT) or inverse NTT, which is the finite field version of the fast Fourier transform (FFT) and takes $O(N\log N)$ time and $O(N)$ space for a degree-$(N-1)$ polynomial.

\paragraph*{Encoding Plaintexts} CKKS supports non-integer messages, so all encoded messages must include a scaling factor $\Delta$. The scaling factor is usually the size of one of the limbs of the ciphertext, which is slightly less than a machine word. 
When multiplying messages together, this scaling factor grows as well. 
The scaling factor must be shrunk down in order to avoid overflowing the ciphertext coefficient modulus.  
We discuss how this procedure works in \Cref{subsec:CKKSSubRout}.

\subsection{Implementing the API}
\label{subsec:CKKSSubRout}
To implement the homomorphic API described in Table~\ref{tab:low-level-api-overview}, we need some ``helper'' subroutines. We first describe these subroutines and then provide the implementations of the homomorphic API using the subroutines.

\paragraph*{Handling a Growing Scaling Factor}
As mentioned in \cref{subsec:CKKSStructure}, all encoded messages in CKKS must have a scaling factor $\Delta$. 
In both the $\PtMult$ and $\Mult$ implementations, the multiplication of the encoded messages results in the product having a scaling factor of $\Delta^2$.
Before these operations can complete, we must shrink the scaling factor back down to $\Delta$ (or at least a value very close to $\Delta$). If this operation is neglected, the scaling factor will eventually grow to overflow the ciphertext modulus, resulting in decryption failure.

To shrink the scaling factor, we divide the ciphertext by $\Delta$ (or a value that is close to $\Delta$) and round the result to the nearest integer. This operation, called $\ModDown$, keeps the scaling factor of the ciphertext roughly the same throughout the computation.\footnote{A better name for this operation would be ``divide and mod-down'' because it reduces the scaling factor {\em as well as} the ciphertext modulus. In this paper, we stick to the standard $\ModDown$ terminology for consistency with the literature.} For a more formal description, we refer the reader to \cite{FullRNSHEAAN}. We sometimes refer to a $\ModDown$ instruction that occurs at the end of an operation as $\Rescale$.

\paragraph*{Handling a Changing Decryption Key}
In both the $\Mult$ and $\Rotate$ implementations, there is an intermediate ciphertext with a decryption key that differs from the decryption key of the input ciphertexts.
In order to change this new decryption key back to the original decryption key, we perform a $\KeySwitch$ operation.
This operation takes in a switching key $\ksk_{\s \rightarrow \s'}$ and a ciphertext $\dbrack{\m}_{\s}$ that is decryptable under a secret key $\s$.
The output of the $\KeySwitch$ operation is a ciphertext $\dbrack{\m}_{\s'}$ that encrypts the same message but is decryptable under a different key $\s'$.

\paragraph*{Key Switching{\em ~\cite{BV11}}} Since the $\KeySwitch$ operation differs between $\Mult$ and $\Rotate$, we do not define it separately.
Instead, we go a level deeper, and define the subroutines necessary to implement $\KeySwitch$ for each of these operations.
In addition to the $\ModDown$ operation, we use the $\ModUp$ operation, which allows us to add primes to our RNS basis.
We follow the structure of the switching key in the work of Han and Ki~\cite{HK19}, where the switching key, parameterized by a length $\dnum$, is a $2 \times \dnum$ matrix of polynomials.
\begin{align} \label{eq:kskShape}
\ksk = \begin{pmatrix} \a_1 & \a_2 & \ldots & \a_\dnum \\ 
\b_1 & \b_2 & \ldots & \b_\dnum \\ 
\end{pmatrix}
\end{align}
The $\KeySwitch$ operation requires that a polynomial be split into $\dnum$ ``digits," then multiplied with the switching key. We define the function $\Decomp$ that splits a polynomial into $\dnum$ digits as well as a $\KSKInProd$ operation to multiply the $\dnum$ digits by the switching key.

Before proceeding further, we refer the reader to \Cref{tab:ckks_subroutines} where all the 
subroutines described above are defined in more detail. The implementation of the API functions are given in Algorithms~\ref{algo:PtMult}, \ref{algo:Mult} and \ref{algo:Rotate}.
We also give a batched rotation algorithm $\HRotate$ in \Cref{algo:HRotate},
which computes many rotations on the same ciphertext faster than applying $\Rotate$ independently several times. 

\begin{algorithm}
\caption{$\PtMult(\dbrack{\m}, \m') = \dbrack{\m\cdot\m'}$}
\label{algo:PtMult}
\begin{algorithmic}[1]
\State $(\a, \b) := \dbrack{\m}$
\State $(\u, \v) := (\a \cdot (\Delta\cdot\m'), \b \cdot (\Delta\cdot\m'))$\\
\Return $(\ModDown_{\calB, 1}(\u), \ModDown_{\calB, 1}(\v))$ \Comment{$\Rescale$}
\end{algorithmic}
\end{algorithm}

\begin{algorithm}
\caption{$\Mult(\dbrack{\m_1}_\s, \dbrack{\m_2}_\s, \ksk_{\s^2\rightarrow \s} ) = \dbrack{\m_1\cdot \m_2}_\s$}
\label{algo:Mult}
\begin{algorithmic}[1]
\State $(\a_1, \b_1) := \dbrack{\m_1}_\s$
\State $(\a_2, \b_2) := \dbrack{\m_2}_\s$
\State $(\a_3, \b_3, \c_3) := (\a_1\a_2, \a_1\b_2 + \a_2\b_1, \b_1\b_2)$
\State $\vec{\a} := \Decomp_\beta(\a_3)$
\State $\hat{\a}[i] := \ModUp(\vec{\a}[i])$ for $1 \leq i \leq \beta$.
\State $(\hat{\u}, \hat{\v}) := \KSKInProd(\ksk_{\s^2\rightarrow \s}, \hat{\a})$
\State $(\u, \v) := (\ModDown(\hat{\u}), \ModDown(\hat{\v}))$ \label{line:multFirstModDown}
\State $(\a', \b') := (\b_3 + \u, \c_3 + \v)$\label{line:multAdd}\\ 
\Return $(\ModDown_{\calB, 1}(\a'), \ModDown_{\calB, 1}(\b'))$ \Comment{$\Rescale$} \label{line:multSecondModDown}
\end{algorithmic}
\end{algorithm}

\begin{algorithm}
\caption{$\Rotate(\dbrack{\m}_\s, k, \ksk_{\psi_k(\s)\rightarrow \s} ) = \dbrack{\phi_k(\m)}_\s$}
\label{algo:Rotate}
\begin{algorithmic}[1]
\State $(\a, \b) := \dbrack{\m}_\s$
\State $(\a_\rot, \b_\rot) := (\Automorph(\a, k), \Automorph(\b, k))$
\State $\vec{\a_\rot} := \Decomp_\beta(\a_\rot)$  \Comment{$\beta$ digits.}
\State $\hat{\a}[i] := \ModUp(\vec{\a_\rot}[i])$ for $1 \leq i \leq \beta$.
\State $(\hat{\u}, \hat{\v}) := \KSKInProd(\ksk_{\psi_k(\s)\rightarrow \s}, {\hat{\a}})$
\State $(\u, \v) := (\ModDown(\hat{\u}), \ModDown(\hat{\v}))$\\
\Return $(\u, \v + \b_\rot)$
\end{algorithmic}
\end{algorithm}

\begin{algorithm}
\caption{$$\HRotate(\dbrack{\m}_\s, \{k_i, \ksk_{\psi_{k_i}(\s)\rightarrow \s}\}_{i=1}^r ) = \{\dbrack{\phi_{k_i}(\m)}_\s\}_{i=1}^r$$}
\label{algo:HRotate}
\begin{algorithmic}[1]
\State $(\a, \b) := \dbrack{\m}_\s$
\State $\vec{\a} := \Decomp_\beta(\a)$  \Comment{$\beta$ digits.}
\State $\hat{\a}[j] := \ModUp(\vec{\a}[j])$ for $1 \leq j \leq \beta$.
\For{$i$ from $1$ to $r$}
\State $\hat{\a}_\rot := \Automorph(\hat{\a}, k_i)$ for $1 \leq j \leq \beta$
\State $(\hat{\u}, \hat{\v}) := \KSKInProd(\ksk_{\psi_{k_i}(\s)\rightarrow \s}, {\hat{\a}_\rot})$
\State $(\u, \v) := (\ModDown(\hat{\u}), \ModDown(\hat{\v}))$
\State $\b_\rot := \Automorph(\b, k_i)$
\State $\dbrack{\phi_{k_i}(\m)}_\s := (\u, \v + \b_\rot)$
\EndFor\\
\Return $\{\dbrack{\phi_{k_i}(\m)}_\s\}_{i=1}^r$
\end{algorithmic}
\end{algorithm}

\begin{table*}[t]
    \centering
    \caption{CKKS Subroutines: \emph{These subroutines enable the implementation of the CKKS API defined in \Cref{tab:low-level-api-overview}.}}
    \label{tab:ckks_subroutines}
    \begin{tabular}{ l l p{0.4in} p{4in}}
    \toprule
    \textbf{Sub-routine Name} & \textbf{Output} & \textbf{Used-in} & \textbf{Description} \\
    \midrule
    $\ModDown_{\calB, d}([\x]_\calB)$ & $[\x/P + e]_{\calB'}$ & $\PtMult$ $\Mult$ $\Rotate$ & This function takes in a polynomial $\x$ in the coefficient representation, where each coefficient is modulo $Q := \prod_{q \in \calB} q$ and represented in the RNS basis $\calB = \{q_1, \ldots, q_\ell\}$. Assume that $d < \ell$ and let $P := \prod_{i=\ell-d+1}^\ell q_i$ be the product of the last $d$ limbs of $\calB$. Let $\calB' = \{q_1, \ldots, q_{\ell-d}\}$, and note that $Q/P = \prod_{q\in \calB'}q$. The output of this function is a polynomial $[\y]_{\calB'}$ where each coefficient of $\y$ equals the corresponding coefficient of $\x$ divided by $P$ plus some small rounding error.\\
    \midrule
    $\ModUp_{\calB, \calB'}([\x]_\calB)$ & $[\x]_{\calB'}$ & $\Mult$ $\Rotate$ & Takes a polynomial $\x$ where each coefficient is in the basis $\calB$ and outputs the representation of $\x$ where each coefficient is in the basis $\calB'$. $\calB$ could be a subset or superset of $\calB'$, or they could be unrelated. Note that this operation must also be performed in the coefficient representation.\\
    \midrule
    $\Decomp_\beta(\x)$ & $\{\x^{(1)}, \ldots, \x^{(\beta)}\}$ & $\Mult$ $\Rotate$ & Takes in a polynomial $\x$ and a parameter $\dnum$ and splits $\x$ into $\dnum$ digits. If $\x$ has $L$ limbs, each digit of $\x$ has roughly $\alpha := \lceil (L+1)/\dnum\rceil$ limbs.\\
    \midrule
    $\KSKInProd(\ksk, \vec{\x})$ & $(\a, \b)$ & $\Mult$ $\Rotate$ & Takes in a key-switching key $\ksk$ with the structure of \cref{eq:kskShape} and a vector of polynomials $\vec{\x}$ of length $\dnum$. Let $\ksk_1$ be the first row of $\ksk$ and let $\ksk_2$ be the second row of $\ksk$. The output of this operation is two polynomials $\a := \langle \ksk_1, \vec{\x} \rangle$ and $\b := \langle \ksk_2, \vec{\x} \rangle$.\\
    \midrule
    $\Automorph(\x, k)$ & $\psi_k(\x)$ & $\Rotate$ & Takes a vector $\x$ with $N$ elements and an integer $k$ and outputs a permutation $\psi_k(\cdot)$ of the elements. This permutation is an automorphism which is {\em not} simply a rotation; intuitively, the permutation $\psi_k$ of an encoded message will result in the decoded value being permuted by the natural rotation $\phi_k$. \\
    \bottomrule
    \end{tabular}
\end{table*}

\medskip\noindent
\textbf{\em Key Takeaway: The Shrinking Ciphertext Modulus}
A main observation coming out of our description of the homomorphic API is that the ciphertext modulus shrinks for each $\PtMult$ (\cref{algo:PtMult}) and $\Mult$ (\cref{algo:Mult}) operation. This occurs in the $\ModDown$ operations at the end of these functions.
If a ciphertext begins with $L$ limbs, we can only compute a circuit with multiplicative depth $L-1$, since the ciphertext modulus shrinks by a number of limbs equal to the multiplicative depth of the circuit being homomorphically evaluated.
This foreshadows the next section where we present an operation called \emph{bootstrapping}~\cite{Gentry09} that increases the ciphertext modulus.

\subsection{Concrete Costs} \label{subsec:lowLevelConCosts}
We present the hardware cost associated with various functions and subroutines in the FHE API in \Cref{tab:aux-cost} and \Cref{tab:low-level-api-cost}, and discuss the content of the tables briefly. 
To generate these performance numbers, we implement an architectural modeling tool that can perform an in-depth analysis given the number of functional units, cache size, and the memory subsystem parameters.
In addition, our tool allows us to tune nearly all parameters of the algorithm, including $N$, $\dnum$, and the maximum ciphertext modulus for a given security level.  

\medskip\noindent
\textbf{\em Key Takeaway: Low Arithmetic Intensity.}
The key take-away from the tables, in particular \Cref{tab:low-level-api-cost}, is that {\em the arithmetic intensity}, defined as the number of operations per byte transferred from DRAM, of all of the functions in the CKKS API is less than $<1$ Op/byte. 
This means that when the ciphertexts do not fit in memory, {\em any natural application (e.g. logistic regression training, neural network evaluation, bootstrapping, etc.) built using these functions will have performance bounded by the memory bandwidth and not the computation speed}. 

Since our ciphertexts will remain too large to fit in the chip cache, much of this work will focus on improving the arithmetic intensity of CKKS bootstrapping. This translates to progressing further in the bootstrapping algorithm per memory transfer, which, in turn, translates to a faster bootstrapping implementation.

\begin{table*}[t]
    \centering
    \caption{Hardware Cost of Auxiliary Subroutines: \emph{These benchmarks were taken for $\log(N) = 17$, $\ell = 35$, $\dnum = 3$. The \textbf{Total Operations} column counts the number of modular additions and multiplications in the operations, (note that this count for the $\Automorph$ function is zero). \textbf{GOP} stands for Giga operations. The \textbf{Total DRAM Transfers} is the sum of \textbf{DRAM Limb Reads}, \textbf{DRAM Limb Writes}, and \textbf{DRAM Key Reads}, the last of which counts the reads specifically for the switching keys. The $\KSKInProd$ operation has no limb writes because the limbs are immediately used in the next operation, the $\ModDown$. The write is counted in the $\ModDown$ when the limbs are written out in to be read back in \emph{slot-wise} format, as discussed in \Cref{sec:macro-fusion}. The \textbf{Arithmetic Intensity} column defines the number of operations per byte transferred from DRAM.}}
    \label{tab:aux-cost}
    \begin{tabular}{cccccccc}
    \toprule
    \specialcell{\textbf{Sub-routine}\\\textbf{Name}} & \specialcell{\textbf{Total Operations}\\\textbf{(in GOP)}} & \specialcell{\textbf{Total Mults}\\\textbf{(in GOP)}} & \specialcell{\textbf{Total DRAM }\\\textbf{Transfers (in GB)}} & \specialcell{\textbf{DRAM Limb}\\\textbf{Reads (in GB)}} & \specialcell{\textbf{DRAM Limb}\\\textbf{Writes (in GB)}} & \specialcell{\textbf{DRAM Key}\\\textbf{Reads (in GB)}} & \specialcell{\textbf{Arithmetic}\\\textbf{Intensity}\\\textbf{(in Op/byte)}}\\
    \midrule
    $\ModDown$ & $0.3000$   & $0.1288$    & $0.1877$    & $0.1007$   & $0.0870$      & $0$  & $\mathbf{1.59}$ \\
    \midrule
    $\ModUp$ & $0.2847$   & $0.1211$     & $0.1510$    & $0.0629$    & $0.0881$     & $0$ & $\mathbf{1.88}$ \\
    \midrule
    $\Decomp$ & $0.0092$  & $0.0092$   & $0.0734$    & $0.0367$     & $0.0367$      & $0$  & $\mathbf{0.12}$ \\
    \midrule
    $\KSKInProd$ & $0.0629$   & $0.0378$    & $0.4530$   & $0.1510$ & $0$    & $0.3020$ & $\mathbf{0.13}$ \\
    \midrule
    $\Automorph$ & $0$  & $0$  & $0.1468$    & $0.0734$  & $0.0734$  & $0$ & $\mathbf{0}$ \\
    \bottomrule
    \end{tabular}
\end{table*}

\begin{table*}[t]
    \centering
    \caption{Hardware Cost of FHE APIs: \emph{These benchmarks were taken for $\log(N) = 17$, $\ell = 35$, $\dnum = 3$. The number of rotations computed in the $\HRotate$ benchmark is $8$. See the caption of \Cref{tab:aux-cost} for a description of the columns.}}
    \label{tab:low-level-api-cost}
    \begin{tabular}{cccccccc}
    \toprule
    \specialcell{\textbf{Operation}\\\textbf{Name}} & \specialcell{\textbf{Total Operations}\\\textbf{(in GOP)}} & \specialcell{\textbf{Total Mults}\\\textbf{(in GOP)}} & \specialcell{\textbf{Total DRAM }\\\textbf{Transfers(in GB)}} & \specialcell{\textbf{DRAM Limb}\\\textbf{Reads (in GB)}} & \specialcell{\textbf{DRAM Limb}\\\textbf{Writes (in GB)}} & \specialcell{\textbf{DRAM Key}\\\textbf{Reads (in GB)}} & \specialcell{\textbf{Arithmetic Intensity}\\\textbf{(in Op/byte)}}\\
    \midrule
    $\PtAdd$ & $0.00459$  & $0$  & $0.1101$ & $0.0734$  & $0.0367$ & $0$ & $\mathbf{0.04}$ \\
    \midrule
    $\Add$ & $0.0092$ & $0$ & $0.2202$ & $0.1468$ & $0.0734$ & $0$ & $\mathbf{0.04}$ \\
    \midrule
    $\PtMult$ & $0.2747$ & $0.1098$ & $0.3282$ & $0.1835$  & $0.1447$ & $0$ & $\mathbf{0.84}$ \\
    \midrule
    $\Mult$  & $1.8333$ & $0.7826$ & $1.9293$  & $0.9070$  & $0.7203$  & $0.3020$ & $\mathbf{0.95}$ \\
    \midrule
    $\Rotate$ & $1.5310$ & $0.6682$ & $1.5645$ & $0.6501$  & $0.6124$ & $0.3020$ & $\mathbf{0.98}$ \\
    \midrule
    $\Conjugate$ & $1.5310$ & $0.6682$ & $1.5645$ & $0.6501$ & $0.6124$ & $0.3020$ & $\mathbf{0.98}$ \\
    \midrule
    $\HRotate$ & $6.2039$ & $2.7363$  & $8.1411$  & $3.2632$ & $2.4621$  & $2.4159$ & $\mathbf{0.76}$ \\
    \bottomrule
    \end{tabular}
\end{table*}
\section{Fully Homomorphic Encryption: Applications} 
\label{section:FHEApplications}

In this section, we describe how the FHE API from \Cref{section:FHEAPI} can be leveraged to develop applications. 
As discussed in \Cref{subsec:CKKSSubRout}, a CKKS ciphertext can only support computation up to a fixed multiplicative depth due to the shrinking ciphertext modulus. 
Once this depth is reached, a \emph{bootstrapping} operation must be performed to grow the ciphertext modulus, which allows for computation to continue. 

Many applications of interest have a deep circuit that requires bootstrapping multiple times: in general, machine learning training algorithms are good examples where deeper circuits for the training computation often lead to greater accuracy of the resulting model. 
In this section, we use \emph{logistic regression training} over encrypted data as a running example to explain the process of FHE-based machine learning training. 
Logistic regression training contains both linear (e.g. inner-products) and non-linear (e.g. sigmoid) operations.
The CKKS scheme naturally supports linear operations, while for non-linear operations we need to use a polynomial approximation (as in \cite{Kim2018LogisticRM,HELogReg}).
The greater the degree of the polynomial, the greater the accuracy of the approximation, which further drives an increase in the circuit depth,  in turn requiring bootstrapping.

\begin{figure}[ht]
  \begin{center}
    \includegraphics[width=1.00\columnwidth]{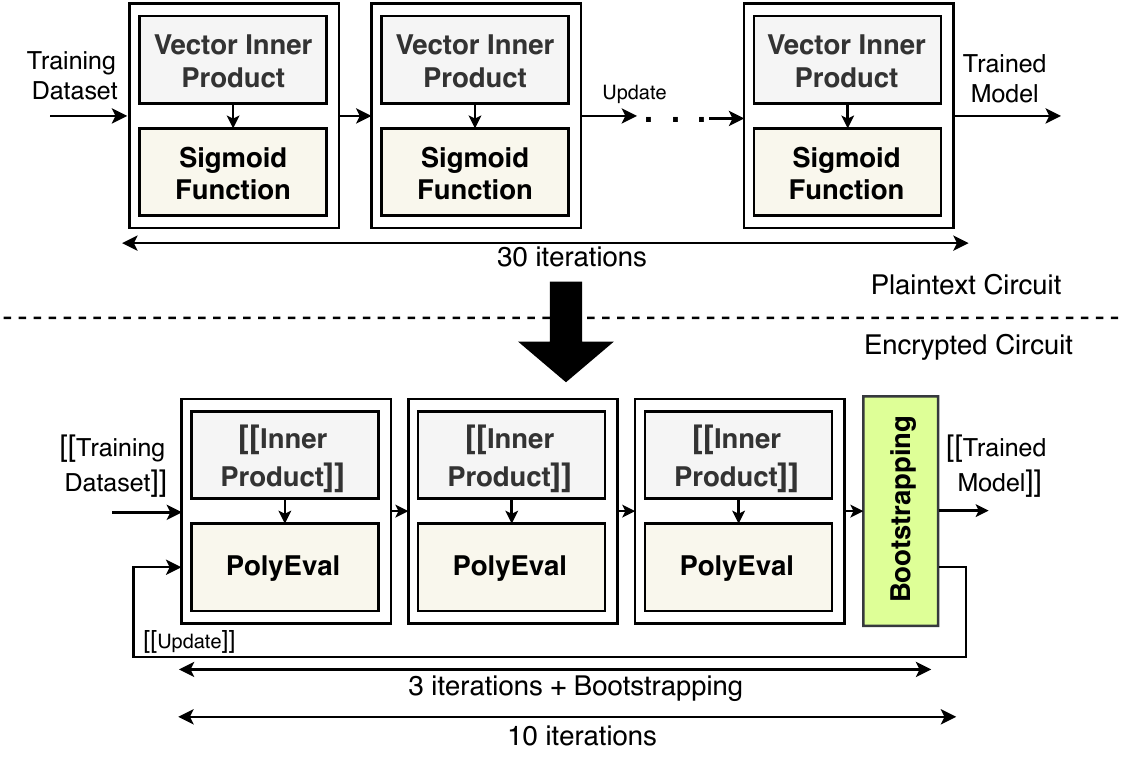} 
  \end{center}
  \vspace{-0.15in}
  \caption{Logistic regression training on encrypted data.}
  \vspace{-0.10in}
  \label{fig:LogRegTraining}
\end{figure}

For our running example, we use the logistic regression training application given in Han, Song, Cheon and Park~\cite{HELogReg} and depicted in \cref{fig:LogRegTraining}. 
The training process is an iterative process that repeatedly computes an inner product followed by a sigmoid function on a training data set and the model weights. 
The logistic regression update equation is as follows.
\begin{align}\label{eq:logRegUpdate}
    \w \gets \w + \frac{\mathsf{lr}}{n}\sum_{i=1}^n \sigma\left(\z_i^T\cdot\w\right) \cdot \z_i
\end{align}
The vector $\w$ is the weight vector, the values $n$ and $\mathsf{lr}$ are scalars, and $\z_i$ represents the $i^{th}$ vector of the training data set.
The $\sigma$ function is the sigmoid function. 

To implement this iterative update, we split the update into two phases: a linear phase that contains the inner product\footnote{In the real implementation of \Cref{eq:logRegUpdate}, these inner products are batched into a matrix-vector product. We use the same algorithm as \cite{HELogReg}.} 
and a non-linear phase that contains the sigmoid function.
We implement these phases separately with common building blocks shown in \Cref{tab:building_blocks}. 
The linear phase can be implemented with an $\InnerProduct$ routine that computes the inner product of two encrypted vectors. 
The non-linear phase is approximated with a polynomial, and the homomorphic evaluation of this polynomial can be implemented with $\PolyEval$. 
The scalar products and summation can be implemented with the $\PtMult$, $\Mult$, and $\Add$ functions. 
After some number of iterations, the encrypted weights are passed through the $\Bootstrap$ routine. 
The exact placement of the $\Bootstrap$ operation in a circuit is application-dependent.
In our running example, bootstrapping needs to be done every three iterations (see Figure~\ref{fig:LogRegTraining}).

\begin{table*}[ht]
    \centering
    \caption{Homomorphic Encryption Application Building Blocks: \emph{These building blocks are implemented using the API from \Cref{tab:low-level-api-overview}.}}
    \label{tab:building_blocks}
    \begin{tabular}{c c p{5in}} \toprule
        \textbf{Name} & \textbf{Output} & \textbf{Description}\\
        \midrule
        $\InnerProduct(\dbrack{\x}, \dbrack{\y})$ & $\dbrack{\langle \x, \y \rangle}$ & Computes the inner product of two encrypted vectors. This computation is the specific encrypted inner product algorithm from Han et al.~\cite{HELogReg}.
        \\
        \midrule
        $\PolyEval(\dbrack{\x}, p(\cdot))$ & $\dbrack{p(\x)}$ & This operation takes an encrypted vector $\x$ and a (univariate) polynomial $p$ as input. The result is an encryption of the evaluation of $p$ at $\x$, where each entry of $p(\x)$ is the evaluation of $p$ on the corresponding entry of $\x$.  \\
        \midrule
        $\PtMatVecMult(\M, \dbrack{\x})$ & $\dbrack{\M\x}$ & This operation takes a plaintext matrix $\M$ and multiplies it by an encrypted vector $\x$. The result is an encryption of the vector $\M\x$. This is a major subroutine in $\Bootstrap$.\\
        \midrule
        $\Bootstrap(\dbrack{\x})$ & $\dbrack{\x}$ & This operation takes in an encryption of a vector $\x$ and outputs an encryption of the same vector $\x$. This operation is necessary to be able to compute indefinitely on encrypted data. Far from being a null operation, this is nearly always the bottleneck operation when computing over encrypted data. \\
        \bottomrule
    \end{tabular}
\end{table*}

\subsection{Bootstrapping}
\label{sec:bootstrapping}

As discussed in \Cref{subsec:CKKSSubRout}, the ciphertext modulus of CKKS shrinks with each multiplication.
In order to compute indefinitely on a CKKS ciphertext, we must grow the ciphertext modulus without also growing the noise. 
This is not as simple as performing a $\ModUp$ function. 
The CKKS bootstrapping procedure~\cite{CKKS20} begins with this $\ModUp$ operation, which gives the new plaintext as $\Delta\cdot \m + \k q$ where $q$ is the modulus for the input ciphertext and $\k$ is some polynomial with small integer coefficients.
The primary goal of the bootstrapping operation is to homomorphically evaluate the modular reduction operation modulo $q$ on this plaintext, returning the plaintext back to $\Delta\cdot \m$. 

The CKKS bootstrapping algorithm follows a general structure that has remained relatively static in the literature~\cite{CCS18, CHH18, HK19, BMTH20, CKKS20} over the past few years. 
This structure has three main components: a linear operation, an approximation of the modular reduction function followed by another linear operation. 
The linear operations in bootstrapping require homomorphically evaluating the DFT on the encrypted data so that we perform modulus reduction on the {\em coefficient representation} of plaintext, rather than the {\em evaluation (or slot) representation}. 
The first of these DFT operations is called $\CoeffToSlot$ and the second is called $\SlotToCoeff$.
In between these two DFT operations is an approximation of the modular reduction function that consists of a polynomial evaluation followed by an exponentiation. 
For further details on polynomial evaluation and the exponentiation, we refer the readers to \cite{HK19, BMTH20}.

To homomorphically evaluate the DFT, we use the observation that the DFT matrix can be factored into submatrices of smaller dimension. 
This turns the homomorphic DFT into a series of $\PtMatVecMult$ operations. 
However, there is a trade-off between the number of $\PtMatVecMult$ operations that must be computed and the size of the matrices in each $\PtMatVecMult$ instance.
Each $\PtMatVecMult$ has a multiplicative depth of 1.
The total dimension of the DFT is $n = N/2 = 2^{16}$ for our parameters. 
Options to evaluate this DFT include evaluating a single $\PtMatVecMult$ with an $n\times n$ input, which would require a very large number of rotations, or evaluating $16$ $\PtMatVecMult$ instances in sequence with only two rotations per instance. 
The former corresponds to treating DFT as a matrix-vector multiplication without using the structure of the DFT matrix while the latter corresponds to running the $O(N\log N)$ algorithm for DFT.

We can interpolate between these two extremes to find the optimal depth vs. computation trade-off. 
Each sub-matrix in the factorization of the DFT matrix has a \emph{radix} corresponding to the number of non-zero diagonals. 
The smaller the radix, fewer the rotations that must be computed during the $\PtMatVecMult$ instance.
The rule is that the product of the radices of the $\PtMatVecMult$ iterations (in the DFT algorithm) must equal $n$. 
For example, for our parameter of $n = 2^{16}$, this gives the options of three $\PtMatVecMult$ iterations with radices of $2^5$, $2^5$, and $2^6$, or five $\PtMatVecMult$ iterations with four iterations having a radix of $2^3$ matrix and one iteration with a radix of $2^4$. 
We call the number of iterations as $\fftIter$.
The homomorphic inverse DFT is computed in an analogous way.

Our approximation of the modular reduction function follows the literature, where we represent the modular reduction function modulo $q$ with a sine function with period $q$, then approximate this sine function with a polynomial.
We represent this polynomial with $\sine(\cdot)$, and we use the Chebyshev polynomial construction used in Han and Ki~\cite{HK19}. 
The degree of this polynomial is $63$.
We give a high-level pseudocode for the bootstrapping algorithm in \Cref{algo:BootstrapHighLevel}.

\begin{algorithm}
\caption{$\Bootstrap(\dbrack{\x}) = \dbrack{\x}$}
\label{algo:BootstrapHighLevel}
\begin{algorithmic}[1]
\State $(\a, \b) := \dbrack{\x}$
\State $\dbrack{\t} := \ModUp(\a, \b)$ \label{line:BootModRaise}
\For{$i$ from $1$ to $\fftIter$}  \Comment{$\CoeffToSlot$ phase.}
\State $\dbrack{\t} \gets \PtMatVecMult(\M_i, \dbrack{\t})$
\EndFor
\State $\dbrack{\t} \gets \PolyEval(\dbrack{\t}, \sine(\cdot))$
\For{$i$ from $1$ to $\fftIter$} \Comment{$\SlotToCoeff$ phase.}
\State $\dbrack{\t} \gets \PtMatVecMult(\M_i, \dbrack{\t})$
\EndFor\\
\Return $\dbrack{\t}$
\end{algorithmic}
\end{algorithm}

\subsection{Concrete Costs}
\label{subsec:applConcCosts}

We give the concrete costs of the logistic regression and $\Bootstrap$ subroutines in \Cref{tab:building-block-cost} and \Cref{tab:boot-cost} respectively. 
As the table shows, the arithmetic intensity of the sub-routines is less than $1$ Op/byte.
\vinod{There has to be a discussion, potentially brief but early in the paper, as to why you are focusing on arithmetic intensity as a measure of performance bottleneck. It is currently taken for granted and I don't think it's that obvious. What if you could reduce the compute and memory by the same amount, that would keep arithmetic intensity the same, but will result in a far more practical scheme?} 
\leo{definitely. the arithmetic intensity discussion is relevant because the ciphertexts cannot fit on memory. I have highlighted this here} 
\vinod{What I'd like is a robust discussion on why we focus on arithmetic intensity as a measure, early on in the paper. I still don't see it.}
As discussed in \Cref{subsec:lowLevelConCosts}, since our ciphertexts do not fit in cache, this means that the performance of all sub-routines is bounded by the main memory bandwidth. 
In \Cref{tab:building-block-cost}, we give benchmarks for the logistic regression implementation based on our architecture modeling discussed in \Cref{subsec:lowLevelConCosts}. 
The parameters we use are from the work of Jung et al.~\cite{GPUBoot21}, and these parameters were chosen to optimize their secure logistic regression application that leverages a GPU implementation of CKKS bootstrapping.
We refer to the original work of Han et al.~\cite{HELogReg} for the full algorithm benchmarked in \Cref{tab:building-block-cost}. 
We note that the logistic regression iteration is the ``most expensive" of the three iterations that follow a $\Bootstrap$, since the ciphertexts in this iteration are the largest. 
As the ciphertext shrinks due to the reduced ciphertext modulus, the computation becomes cheaper. 
However, the arithmetic intensity remains essentially the same, 
and the performance of each phase of the algorithm is bottle-necked by the memory bandwidth. 
Overall, roughly half of the total runtime is spent in bootstrapping.

\medskip\noindent
\textbf{\em Key Takeaway:}
Bootstrapping is often the bottle-neck operation in HE applications, especially applications that implement a deep circuit. 
For example, even when using a heavily-optimized GPU implementation of bootstrapping, nearly half of the time in HE logistic regression training is spent on bootstrapping~\cite{GPUBoot21} (\cref{tab:building-block-cost}). 
This motivates the need to optimize the $\Bootstrap$ operation to efficiently support deep circuits.
Furthermore, the building blocks of bootstrapping are the same as many other HE applications; there are essentially no subroutines that are unique to bootstrapping. 
Many of the optimizations we give in \Cref{sec:cachingopts} and \Cref{sec:algoopts} apply more generally to HE applications. 

\begin{table*}[ht]
    \centering
    \caption{Hardware Cost of FHE Applications: \emph{These benchmarks were taken for $\log(N) = 17$, $\ell = 35$, $\dnum = 3$. See the caption of \Cref{tab:aux-cost} for a description of the columns. The number of features in the logistic regression is $d = 256$. The $\InnerProduct$ and $\PolyEval$ benchmarks are for the first iterations after a $\Bootstrap$. The ``Full LR Iteration'' row is the first iteration of the training algorithm after a $\Bootstrap$. The degree of the polynomial evaluated in $\PolyEval$ is $3$. 
    }}
    \label{tab:building-block-cost}
    \begin{tabular}{cccccccc}
    \toprule
    \specialcell{\textbf{Sub-routine}\\\textbf{Name}} & \specialcell{\textbf{Total Operations}\\\textbf{(in GOP)}} & \specialcell{\textbf{Total Mults}\\\textbf{(in GOP)}} & \specialcell{\textbf{Total DRAM }\\\textbf{Transfers(in GB)}} & \specialcell{\textbf{DRAM Limb}\\\textbf{Reads (in GB)}} & \specialcell{\textbf{DRAM Limb}\\\textbf{Writes (in GB)}} & \specialcell{\textbf{DRAM Key}\\\textbf{Reads (in GB)}} & \specialcell{\textbf{Arithmetic}\\\textbf{Intensity}\\\textbf{(in Op/byte)}}\\
    \midrule
    $\InnerProduct$ & $7.8558$  & $3.3806$ & $16.5413$ & $7.2918$  & $4.8455$ & $4.4040$ & $\mathbf{0.47}$\\
    \midrule
    $\PolyEval$ & $2.9314$ & $1.2188$ & $3.5484$ & $1.7144$ & $1.3118$ & $0.5222$ & $\mathbf{0.83}$ \\
    \midrule
    Full LR Iteration & $92.4225$ & $39.6322$ & $195.052$ & $86.7822$  & $56.1387$ & $52.131$ & $\mathbf{0.47}$ \\
    \midrule
    $\Bootstrap$ & $149.546$ & $64.6859$  & $207.982$  & $109.91$ & $65.2434$  & $32.8288$ & $\mathbf{0.72}$ \\
    \bottomrule
    \end{tabular}
\end{table*}

\begin{table*}[ht]
    \centering
    \caption{Hardware Cost of Bootstrapping: \emph{These benchmarks were taken for $\log(N) = 17$, $\ell = 35$, $\dnum = 3$. See the caption of \Cref{tab:aux-cost} for a description of the columns. These benchmarks represent the performance of the main sub-routines of bootstrapping. The degree of the polynomial in $\PolyEval$ is $63$.}}
    \label{tab:boot-cost}
    \begin{tabular}{cccccccc}
    \toprule
    \specialcell{\textbf{Sub-routine}\\\textbf{Name}} & \specialcell{\textbf{Total Operations}\\\textbf{(in GOP)}} & \specialcell{\textbf{Total Mults}\\\textbf{(in GOP)}} & \specialcell{\textbf{Total DRAM }\\\textbf{Transfers(in GB)}} & \specialcell{\textbf{DRAM Limb}\\\textbf{Reads (in GB)}} & \specialcell{\textbf{DRAM Limb}\\\textbf{Writes (in GB)}} & \specialcell{\textbf{DRAM Key}\\\textbf{Reads (in GB)}} & \specialcell{\textbf{Arithmetic}\\\textbf{Intensity}\\\textbf{(in Op/byte)}}\\
    \midrule
    $\CoeffToSlot$ & $58.486$   & $25.8087$  & $86.7424$ & $46.8651$ & $25.2875$ & $14.5899$ & $\mathbf{0.67}$ \\
    \midrule
    $\PolyEval$ & $57.834$ & $24.4496$ & $65.643$ & $33.0406$ & $23.744$ & $8.8584$ & $\mathbf{0.88}$ \\
    \midrule
    $\SlotToCoeff$ & $33.2265$ & $14.4275$  & $55.5001$  & $30.004$  & $16.1156$ & $9.3806$ & $\mathbf{0.59}$ \\
    \bottomrule
    \end{tabular}
\end{table*}
\section{CKKS Bootstrapping: Caching Optimizations}
\label{sec:cachingopts}

In this section and \cref{sec:algoopts}, we present our optimizations to the CKKS bootstrapping algorithm.
These optimizations fall into two categories: those that rely on hardware assumptions and those that do not.
Our first class of optimizations assume a lower bound on the amount of available cache size relative to the size of the ciphertext limbs while second class of optimizations are more general as they reduce the total operation count of CKKS bootstrapping as well as the total number of DRAM reads, regardless of the hardware architecture. 

This section focuses on the first set of optimizations.
These caching optimizations do not affect the operation count of $\Bootstrap$; instead, they reduce DRAM reads and writes to reduce the overall memory bandwidth requirement.
Our optimizations demonstrate how best to utilize caches of various sizes relative to the size of the ciphertext limbs.
We quantify the improvements of these optimizations in \Cref{sec:cachingoptsTakeaway}, where we give benchmarks for progressively larger cache sizes.
Our baseline benchmark is the parameter set from the GPU bootstrapping implementation of Jung et al.~\cite{GPUBoot21}.
The parameters are given in \Cref{tab:bootParams}.

\subsection{Caching $O(1)$ Limbs} 
\label{sec:macro-fusion}
This is the first in a series of optimizations that details how best to utilize a cache for various cache sizes relative to the ciphertext limbs.
We begin by discussing how to utilize a cache that can store a constant number of limbs.
Intuitively, this optimization computes as much as possible on a single limb before writing it back to the main memory.
This often involves performing the operations of several higher-level functions on a single limb before beginning the same sequence of operations on the next limb.
This technique was referred to by Jung et al.~\cite{GPUBoot21} as a ``fusing" of operations, and we include all fusing operations listed in their work in our bootstrapping algorithm. 
In addition, we provide a novel data mapping technique to handle caching data with different \emph{data access patterns}.

\paragraph*{Data Access Patterns}
Having a small-cache (about $1$-$3$~MB) in any FHE compute system has a caveat that must be carefully addressed. 
Some operations in CKKS such as $\NTT$ and $\iNTT$ operate on data within the slots of the same limb, independent of the other limbs in the ciphertext.
On the other hand, RNS basis change operations in $\ModUp$ and $\ModDown$ require interaction between a certain number of slots across various limbs.
This requires having a few slots from multiple limbs in on-chip memory to reduce the number of accesses to main memory for a single operation.
To account for this, we define two different types of data access patterns.
For the functions where limbs can be operated upon independently, we define the data access pattern as \emph{limb-wise} and for the functions where slots can be operated upon independently, we define the data access pattern as \emph{slot-wise}. 
A summary of this is given in \Cref{tab:data_access}.
We have also illustrated this by giving a high-level pseudo code of $\ModUp$ in \Cref{algo:ModUp}. 
From this algorithm, it is evident that the $\ModUp$ operation includes both \emph{limb-wise} and \emph{slot-wise} operations, requiring a memory mapping that is efficient for both access patterns.
A naive memory mapping would result in low throughput for at least one of these access patterns.
Therefore, we describe a novel memory mapping approach to handle these two access patterns.

\begin{algorithm}
\caption{$\ModUp_{\calB, \calB\cup\calB'}([\x]_\calB) = [\x]_{\calB \cup \calB'}$}
\label{algo:ModUp}
\begin{algorithmic}[1]
\For{$i$ from $1$ to $|\calB|$}
\State $[\x]_i \gets \iNTT([\x]_i)$ \Comment{\emph{limb-wise}}
\EndFor
\For{$j$ from $1$ to $|\calB'|$} \Comment{Basis conversion.}
\State $[\x]_j \gets \NewLimb_j([\x]_1, \ldots, [\x]_{|\calB|})$ \label{line:modUpBasisConv} \Comment{\emph{slot-wise}}
\EndFor
\For{$j$ from $1$ to $|\calB'|$}
\State $[\x]_j \gets \NTT([\x]_j)$   \Comment{\emph{limb-wise}}
\EndFor\\
\Return $[\x]_{\calB \cup \calB'}$
\end{algorithmic}
\end{algorithm}

\begin{table}[ht]
    \centering
    \caption{Data dependencies and access pattern in Different Functions 
    \\\emph{The $\NewLimb$ function is used in both $\ModUp$ and $\ModDown$.}
    } 
    \begin{tabular}{cccc} \toprule 
        \textbf{Operation} & \textbf{Interaction} & \textbf{Independent} & \textbf{Access pattern} \\ \midrule 
        $\NTT$, $\iNTT$ & Intra-limb & Inter-limb & \emph{limb-wise} \\ 
        $\NewLimb$ & Inter-limb & Intra-limb & \emph{slot-wise} \\
        \bottomrule
    \end{tabular}
    \label{tab:data_access}
    \vspace{-0.12in}
\end{table}

\begin{figure}
   \begin{center}
     \includegraphics[width=1.00\columnwidth]{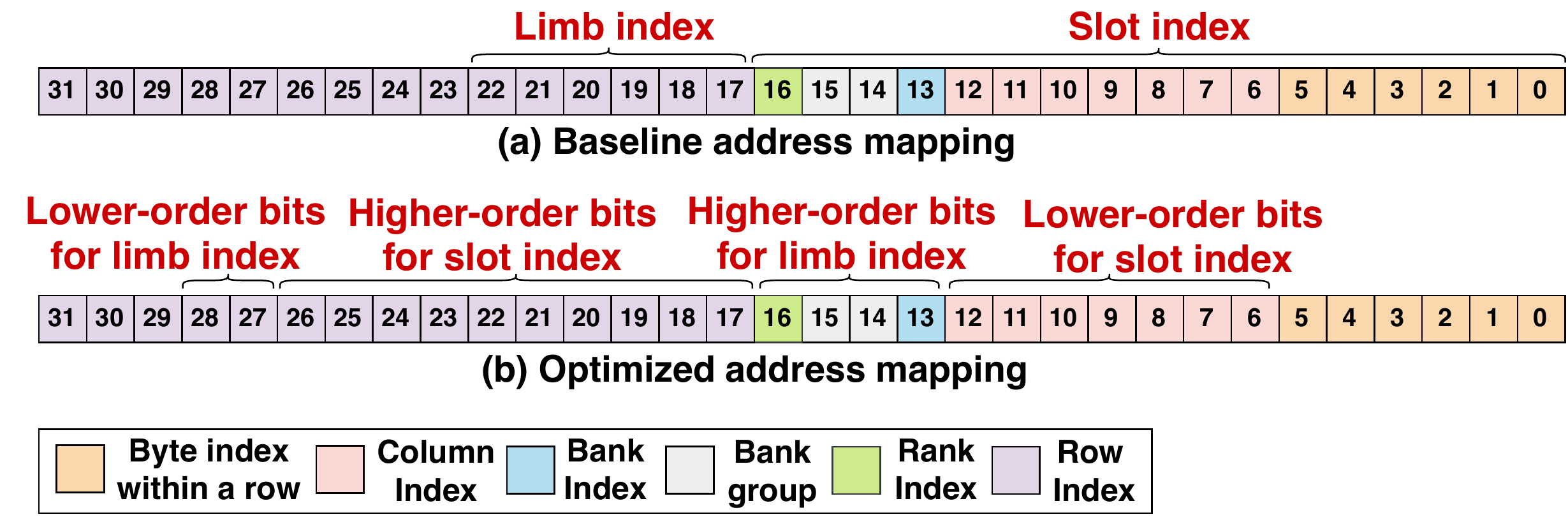} 
   \end{center}
   \vspace{-0.15in}
   \caption{DDR4 physical address mapping. \emph{Baseline address mapping indexes all the slots ($2^{17}$) using the lower-order $17$ bits and all limbs using the immediate next $6$ bits. In optimized physical address mapping, slots are indexed using $7$ bits from the column and $10$ bits from the row, accounting for $2^{17}$ slots. The limbs are indexed using the $4$ bits that index bank group, bank, and rank and $2$ bits from the row index.}}
   \vspace{-0.12in}
   \label{fig:AddrMapping}
 \end{figure}

\paragraph*{Physical Address Mapping}
When we re-purpose the last level cache to support both \emph{limb-wise} and \emph{slot-wise} access patterns, we observe that the physical address mapping of the data in main memory has a substantial impact on the time it takes to transfer data from the main memory.  
Figure~\ref{fig:AddrMapping} (a) shows a natural physical address mapping for a ciphertext.
We call this the baseline address mapping.
Through simulations in DRAMSim3~\cite{LYRSJ20} we notice that for this baseline address mapping, the \emph{limb-wise} accesses require $2.3$~ms to read $35$~limbs worth of data.
However, we notice that the \emph{slot-wise} access pattern requires $9.2$~ms to transfer the same amount of data.
This is significantly lower as with the peak theoretical bandwidth (i.e., $19.2$~GB/s) for DDR4 the time required to read $35$~limbs worth of data is $1.9$~ms.

There are two reasons for this performance hit while doing \emph{slot-wise} accesses. 
With $L=35$, the size of the ciphertext is $36.7$~MB whose limbs can be stored sequentially within a memory bank in one of the bank groups in main memory.
Each limb of the ciphertext spans across multiple rows of the memory bank.
Typically, each bank in main memory has a currently activated row whose contents are copied into a row buffer (acting as a cache) that can be accessed quickly. 
However, with \emph{slot-wise} access pattern, every access is trying to read a different row, which takes longer because each row must be activated first.
Moreover, with \emph{slot-wise} accesses, we are unable to exploit the fact that bank accesses to different banks' groups require less time delay between accesses in comparison to the bank accesses within the same bank's group.   
Instead, we keep accessing data from the memory bank within the same bank group.

We propose an optimized physical address mapping as shown in Figure~\ref{fig:AddrMapping} (b).
As shown in \Cref{tab:read_times}, with this proposed address mapping, we observe that the \emph{limb-wise} access requires a data transfer time of $2.5$~ms, which is about $8\%$ reduction in the times observed for the baseline \emph{limb-wise} accesses.
However, compared to the baseline \emph{slot-wise} access pattern, our optimized \emph{slot-wise} access pattern sees an increase in data transfer time by $76\%$, which is a significant improvement.
We observe that the total data transfer time for baseline address mapping is about $2.4\times$ higher than our optimized mapping. 
Our optimized physical address mapping ensures that when performing \emph{limb-wise} and \emph{slot-wise} reads/writes, we exploit bank-level parallelism, and we focus on reducing the bank thrashing by not changing a bank's currently activated row frequently. 
Note that for a different DRAM type such as HBM2 or GDDR5/6, similar physical address mappings can be done to optimize the main memory bandwidth utilization.  

\begin{table}[ht]
    \centering
    \caption{DRAM transfer times with Baseline and Optimized mapping for different access patterns: \emph{Transfer times are computed for reading $L=35$ limbs worth of data, which is $36.7$~MB for our baseline parameter set.}} 
    \begin{tabular}{cccc} \toprule 
        \textbf{Mapping} & \specialcell{\textbf{\emph{limb-wise}}\\\textbf{access}} & \specialcell{\textbf{\emph{slot-wise}}\\\textbf{access}} & 
        \textbf{Total Time} \\ \midrule 
        Baseline & $2.3$~ms & $9.2$~ms & $11.5$~ms \\
        Optimized & $2.5$~ms &  $2.2$~ms & $4.7$~ms \\ 
        \bottomrule
    \end{tabular}
    \label{tab:read_times}
    \vspace{-0.12in}
\end{table}

\subsection{$\beta$-Limb Caching}
\label{sec:beta-limb-caching}
The next optimization considers a cache size that is $O(\beta)$. Recall that $\beta$ is the number of digits generated from a polynomial key switching. We refer to Han and Ki~\cite{HK19} for more details. For our parameters where $\beta \leq \dnum = 3$, this amounts to about $6$~MB of cache.
We need space for $3$ limbs at all-times and $3$ limbs worth of space to store intermediate results and other required constants. 
With this optimization, we can greatly reduce the number of accesses to main memory during key-switching.

Consider the $\HRotate$ function in \Cref{algo:HRotate}. 
There are $\beta$ digits that are produced as the output of the $\ModUp$ operations. 
Naively, for each rotation we would read the limbs for each of the $\beta$ digits, rotate them, then compute the inner product with the key-switching key. 
Since now we have space in the cache for $\beta$ digits, we can instead pull in a single limb from each of the $\beta$ outputs of $\ModUp$, then compute the rotation and the inner product with the switching key limbs all at once.
This allows us to read in the outputs of the $\ModUp$ function only once, regardless of the number of rotations computed. 

\subsection{$\alpha$-Limb Caching}
\label{sec:alpha-limb-caching}

For this optimization, we assume that we have a relatively large LLC that can hold $O(\alpha)$ limbs.
Recall that $\alpha$ is the number of limbs in a single digit after output by the $\Decomp$ function for key switching. We refer to Han and Ki~\cite{HK19} for more details. 
In practice, this optimization requires only slightly more than $2\alpha$ limbs, using about $27$~MB ($2\alpha$ + $3$~MB) for $\alpha=\lceil L + 1/\dnum \rceil = 12$ as $L=35$ and $\dnum=3$.

Under this assumption, we observe a dramatic decrease in the number of accesses to the main memory.
This is because all of the \emph{slot-wise} basis conversion operations in $\ModUp$ (line~\ref{line:modUpBasisConv} in \cref{algo:ModUp}) and $\ModDown$ operate over $\alpha$ limbs.
If we can fit these $\alpha$ limbs in cache, then we can generate new limbs in their entirety within the cache.
With each new limb in cache, we can perform the NTT on the limb, which completes the basis change operation, and write this limb out to memory.
This lets us generate all new limbs in evaluation format without having to write them out in \emph{slot-wise} format and then reading them back in \emph{limb-wise} format. 

\paragraph*{Accumulator Caching} We briefly mention an optimization that is easily enabled by a large cache but is also available with smaller caches ($O(\beta)$ or even smaller).
This optimization improves the memory bandwidth of the baby-step giant-step polynomial evaluation from Han and Ki~\cite{HK19}.
A straight-forward optimization is to cache the leaves (the baby-step) \vinod{What is this? A full understanding of this requires understanding what bsgs is! so you have to describe it. maybe in an appendix.} \vinod{Please add this, but you can do it in the next draft.} polynomials and reuse them to compute all of the giant-step limbs.
However, if there is not enough space for the baby-step polynomials, we can still save DRAM reads by caching the partial sums of the giant step limb.
When we read in a baby-step limb, we add this limb to all cached accumulators.

\subsection{Re-Ordering Limb Computations} 
\label{sec:reorder}
For the $\ModDown$ operation, the limbs that are being reduced need additional operations to be performed on them.
The $\ModDown$ operations in key switching and bootstrapping drop $\alpha$ limbs. 
In this re-ordering optimization, we propose computing these $\alpha$ limbs first so that the additional operations can be performed immediately.
This optimization is especially potent when these $\alpha$ limbs can be cached, since then there is no need to write out these limbs as they are being computed. 
Once we have the $\alpha$ limbs, we can begin the $\ModDown$ operation by computing the output of the basis conversion.
Then, for each subsequent limb that is computed, this limb can be immediately combined with the basis conversion output, saving DRAM transfers.

\subsection{Key Takeaway} 
\label{sec:cachingoptsTakeaway}
\begin{figure}[t]
  \begin{center}
     \includegraphics[width=\columnwidth]{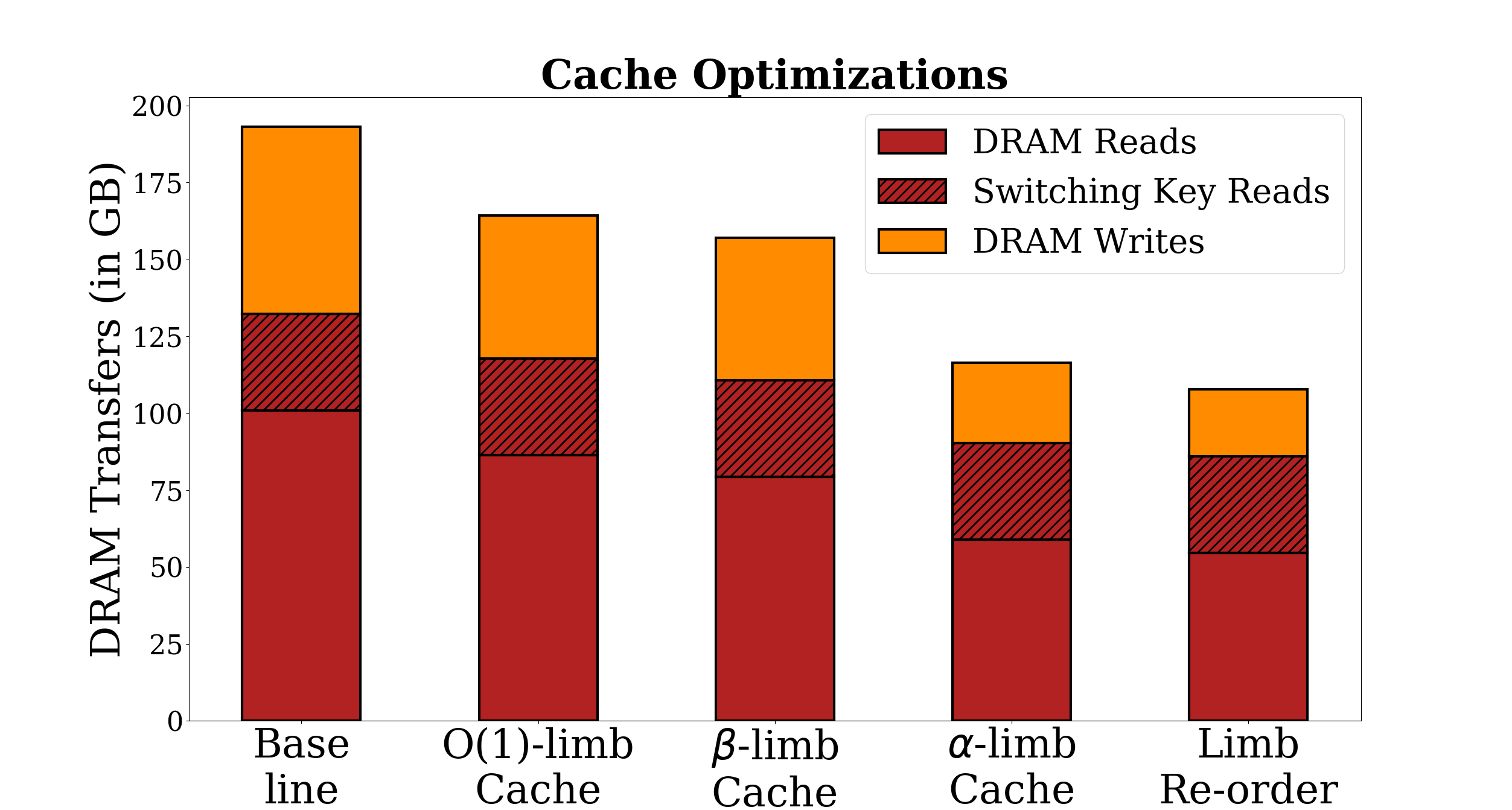} 
  \end{center}
  \vspace{-0.15in}
  \caption{DRAM transfers with various memory optimizations. As the cache size grows from left-to-right more optimizations become available. The impact is assessed cumulatively i.e. each successive optimization builds on top of the earlier ones. The order of the optimizations correspond to the order of the sections in \Cref{sec:cachingopts}. 
  }
  \vspace{-0.15in}
  \label{fig:cacheBarPlot}
 \end{figure}

The benefits of the optimizations in this section are presented in \Cref{fig:cacheBarPlot}. 
As the figure shows, growing the cache size reduces the DRAM transfers of the bootstrapping algorithm by employing the optimizations described in this section. 
Note that the number of compute operations in the bootstrapping algorithm remains fixed for all these benchmarks. 

\section{CKKS Bootstrapping: Algorithmic Optimizations} 
\label{sec:algoopts}
In this section, we present our algorithmic optimizations to the CKKS bootstrapping algorithm. 
These optimizations represent strict improvements to the CKKS bootstrapping algorithm and they do not depend on the cache size.
However, as an added benefit of reducing the compute operation count, they also reduce the memory bandwidth, as displayed in \Cref{fig:algoBarPlot}. 

Our baseline for demonstrating the improvements of these optimizations is the memory-optimized algorithm from \Cref{sec:cachingopts}.
Therefore, the left-most baseline bar in \Cref{fig:algoBarPlot} contains all of the memory optimizations described in \Cref{sec:cachingopts}.
For the algorithm that includes all of our optimizations, we performed a parameter search to optimize the bootstrapping throughput for a $128$-bit security level.
We discuss our parameter search method further in \Cref{section:eval}.
These parameters are given in \Cref{tab:bootParams}, and all benchmarks in \Cref{fig:algoBarPlot} were taken using these same parameters.

\subsection{Combining $\ModDown$ and $\Rescale$ in $\Mult$}
\label{sec:moddown-rescale}
This optimization merges the two $\ModDown$ operations in lines~\ref{line:multFirstModDown} and~\ref{line:multSecondModDown} in \Cref{algo:Mult}. 
To merge these $\ModDown$ operations, we must lift the addition step in line~\ref{line:multAdd} above the first $\ModDown$.  
We achieve this by modifying the double-hoisting method from Bossuat et al.~\cite{BMTH20}, multiplying the two polynomials by $P$ to efficiently lift the two polynomial to the modulus $PQ$. 
We denote the operation that multiplies by $P$ modulo $Q$ and then interprets the result modulo $PQ$ as $\PModUp$. 
By applying the $\PModUp$ function, we can move the addition above the first $\ModDown$, making the two $\ModDown$ operations adjacent, which allows them to be combined. 
This new $\Mult$ algorithm, denoted as $\NewMult$, is given in \Cref{algo:NewMult}, and the lines in blue denote the differences from \Cref{algo:Mult}. 

\paragraph*{Faster Encrypted Inner Product}
As a direct result of this optimization, we obtain a faster encrypted inner product. 
Consider the operation that computes $\dbrack{\z} = \sum_i \Mult(\dbrack{\x_i}, \dbrack{\y_i})$ where $\vec{\dbrack{\x}}$ and $\vec{\dbrack{\y}}$ are vectors of ciphertexts. 
Using the $\NewMult$ operation, we need to compute only one $\ModDown$ operation over the entire sum. 
This is because we can merge the additions in line~\ref{line:newMultSum} to sum all of the polynomials before any $\ModDown$ is computed.

\begin{algorithm}
\caption{
$\NewMult(\dbrack{\m_1}_\s, \dbrack{\m_2}_\s, \ksk ) = \dbrack{\m_1\cdot \m_2}_\s$}
\label{algo:NewMult}
\begin{algorithmic}[1]
\State $(\a_1, \b_1) := \dbrack{\m_1}_\s$
\State $(\a_2, \b_2) := \dbrack{\m_2}_\s$
\State $(\a_3, \b_3, \c_3) := (\a_1\a_2, \a_1\b_2 + \a_2\b_1, \b_1\b_2)$
\State $\vec{\a} := \Decomp_\dnum(\a_3)$
\State $\hat{\a}_i := \ModUp(\vec{\a}[i])$ for $1 \leq i \leq \dnum$.
\State $(\hat{\u}, \hat{\v}) := \KSKInProd(\ksk_{\s^2\rightarrow \s}, \hat{\a})$
\State {\color{blue}$(\hat{\b}_3, \hat{\c}_3) := (\PModUp(\b_3), \PModUp(\c_3))$}\\
\Return {\color{blue}$(\ModDown(\hat{\u} + \hat{\b}_3), \ModDown(\hat{\v} + \hat{\c}_3))$} \label{line:newMultSum}
\end{algorithmic}
\end{algorithm}

\subsection{Hoisting the $\ModDown$ in $\PtMatVecMult$}
\label{sec:rotation-hoisting}
In section~\ref{subsec:CKKSSubRout}, we discussed how $r$ rotations on the same ciphertext can be computed more efficiently than simply applying the $\Rotate$ function $r$ times. 
This function $\HRotate$ described in \Cref{algo:HRotate} achieves an improved performance by identifying an expensive common subroutine in all of the $\Rotate$ operations: the $\ModUp$ routine. 

Bossuat et al.~\cite{BMTH20} present an optimization that hoists the second \emph{slot-wise} operation in the function: the $\ModDown$ routine. 
However, their technique is similar to the one in $\NewMult$, where the message polynomial is lifted to the raised modulus via the inexpensive $\PModUp$ procedure. 
They call this optimization ``double-hoisting."
Our $\ModDown$ hoisting optimization is used in the context of a baby-step giant-step (BSGS) algorithm that implements $\PtMatVecMult$. 
The trade-off in this algorithm is that a larger baby-step and a smaller giant step means more DRAM reads for the switching keys, while a smaller baby-step and a larger giant step means more DRAM reads for the ciphertexts, since the baby-step ciphertexts must be read in for each giant-step. 

In \Cref{sec:key-compression}, we give a simple optimization to compress the size of the keys by a factor of $2$. 
Using our architecture modeling tool, we determine that this optimization shifts the balance between the baby-step size and the giant-step size so significantly that the optimal number of giant steps is $1$. 
This essentially collapses the baby-step giant-step structure into just a single step that computes all $r$ iterations at once.
Therefore, by removing the giant steps in the BSGS algorithm, the $\PtMatVecMult$ collapses into a single instance of $\HRotate$ that includes the $\PModUp$ double-hoisting optimization, which allows the $\PtMult$ to be absorbed into the inner loop. 
This algorithm is given in \Cref{algo:PtMatVecMult}, and the lines that differ from $\HRotate$ are in blue.

\begin{algorithm}
\caption{
$\PtMatVecMult(\M, \dbrack{\x}, \{k_i, \ksk_i\}_{i=1}^r ) = \dbrack{\M\x}$}
\label{algo:PtMatVecMult}
\begin{algorithmic}[1]
\State $(\a_\x, \b_\x) := \dbrack{\x}_\s$
\State $\vec{\a_\x} := \Decomp_\beta(\a_\x)$  \Comment{$\beta$ digits.}
\State $\hat{\a}_j := \ModUp(\vec{\a_\x}^{(i)})$ for $1 \leq j \leq \beta$.
\State {\color{blue}$(\hat{\a}_\y, \hat{\b}_\y) \gets 0, 0$ \Comment{We will have $\y = \M\x$.}}
\For{$i$ from $1$ to $r$}
\State $\hat{\a}_\rot^{(j)} := \Automorph(\hat{\a}_j, k_i)$ for $1 \leq j \leq \beta$
\State $(\hat{\u}, \hat{\v}) := \KSKInProd(\ksk_{i}, \vec{\hat{\a}_\rot})$
\State $\b_\rot := \Automorph(\b_\m, k_i)$
\State {\color{blue} $\hat{\b}_\rot, \hat{\M}_i \gets \PModUp(\b_\rot), \PModUp(\Delta \cdot \M_i)$}
\State {\color{blue}\Comment{$\M_i$ is the $i^{th}$ non-zero diagonal of $\M$}}
\State {\color{blue} $(\hat{\a}_\y, \hat{\b}_\y) \mathrel{+}= \hat{\M}_i\cdot(\hat{\u}, \hat{\v} + \hat{\b}_\rot)$ \Comment{$\PtMult$}}
\EndFor\\
\Return {\color{blue}$(\ModDown(\hat{\a}_\y), \ModDown(\hat{\b}_\y))$}
\end{algorithmic}
\end{algorithm}

\paragraph*{Removing Giant-Steps Beyond Bootstrapping}
This optimization is not a bootstrapping-only optimization.
The hoisting optimizations that are described for $\PtMatVecMult$ for bootstrapping are more broadly applicable to the $\InnerProduct$ computation. 
When multiple $\InnerProduct$ operations needs to be performed in parallel, this hoisting optimization can be amortized across these parallel $\InnerProduct$ computations, which results in about $35\%$ improvement in logistic regression training iterations for our running example. 

\subsection{Compressing the Key with a PRNG}
\label{sec:key-compression}
This optimization is not our own; rather, it is a folklore technique often used to reduce communication when sending ciphertexts or keys over a network (e.g. it is used in Kyber, a leading candidate public-key encryption scheme in the ongoing NIST post-quantum cryptography standardization~\cite{BDKLLSSSS18}). 
However, to our knowledge, we are the first to use this optimization to reduce the memory bandwidth for hardware acceleration of homomorphic encryption as well as the first to analyze this optimization alongside the other optimizations listed in this section. 
As discussed in \Cref{sec:rotation-hoisting}, this optimization has subtle yet highly impactful effects on the other optimizations that we list, drastically changing the optimal parameters for CKKS bootstrapping. 

This optimization is a natural result of the observation that half of the switching key consists of truly random polynomials. 
By replacing these truly random polynomials with pseudorandom polynomials generated via PRNG, we can avoid shipping the large random polynomials to and from DRAM, instead sending only the short PRNG key. 

\begin{figure}[t]
  \begin{center}
    \includegraphics[width=1\columnwidth]{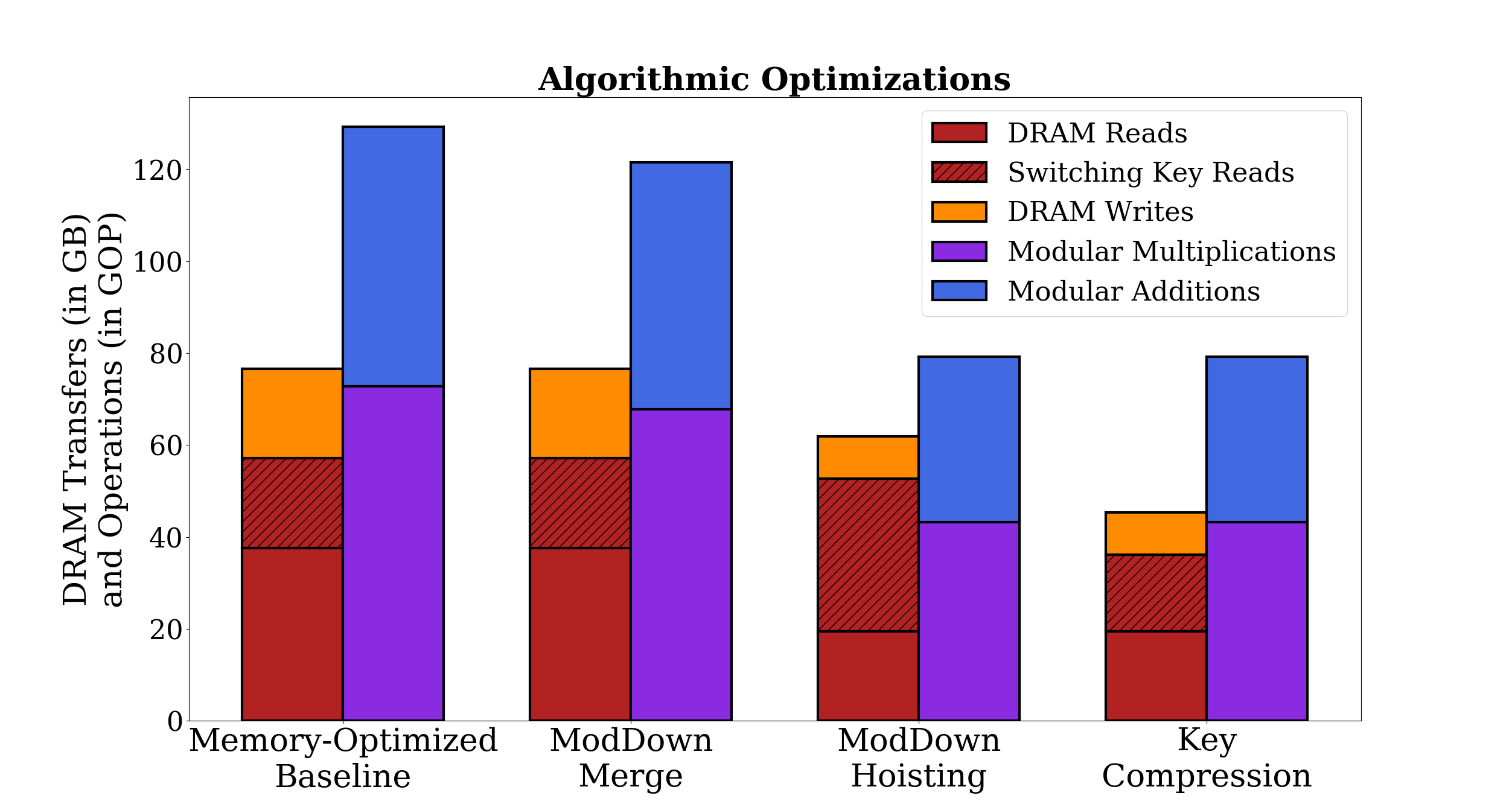}
  \end{center}
  \vspace{-0.15in}
  \caption{
  This figure displays the algorithmic optimizations described in \Cref{sec:algoopts}. The impact is assessed cumulatively i.e. each successive optimization builds on top of the earlier ones. The baseline benchmark begins with all of the memory optimizations from \Cref{sec:cachingopts}. All benchmarks are taken with the \textbf{Best-case Parameters} from \Cref{tab:bootParams}. GOP on y-axis stands for Giga operations.}
  \vspace{-0.10in}
  \label{fig:algoBarPlot}
 \end{figure}

\subsection{Key Takeaways} 
\label{sec:algoOptTakeaway}
Figure~\ref{fig:algoBarPlot} shows how various optimizations impact the operation count and the DRAM transfers for CKKS bootstrapping. 
Moving from left to right on the plot, arithmetic intensity starts to improve as each successive optimization is applied and enabling all our optimizations result in a cumulative $2.43\times$ improvement and a final arithmetic intensity value of $1.75$. 
We now contextualize this compute and bandwidth optimization in the context of current computing platforms.

\paragraph*{Datacenter CPUs}
Consider an example of a top-of-line datacenter CPU such as the AMD EPYC 7763. 
This CPU supports a maximum of $128$ parallel thread across $64$ SMT cores running at a base clock frequency of $2.45$~GHz. 
This configuration supports peak integer theoretical throughput of $2.5$~TOp/s (Each operation here is a $64$-bit Integer Fused Multiply Add in AVX256 mode). 
Each socket consists of $8$ compute die (CCD) with a local $32$~MiB L3 cache per die. 
The total L3 cache per socket comes out to $256$~MiB. 
Additionally, the socket offers an $8$-channel DDR4-$3200$ memory subsystem with an aggregate bandwidth of $204$~GB/s.

At first glance the total L3 capacity appears to be more than sufficient for storing multiple ciphertexts in cache. 
However the die-to-die bandwidth is limited by the underlying interconnect (Infinity Fabric) to $51.2$~GiB/s reads and $25.6$~GiB/s writes. 
There are similar bandwidth limits at the L1-L2 and L2-L3 interfaces on each die. 
Thus, it is necessary to consider the compute available on each die in the context of the bandwidth available to that die.

Each CCD pairs $310$~GOp/s with $51.2$~GiB/s of memory bandwidth.
This gives a theoretical INT64 FMA arithmetic intensity of $\sim 6$. 
On current hardware, $64$-bit modular operations need to be emulated using multiple arithmetic operations as seen in section \ref{ssec:mod_arith}.
Compensating for this, we observe that the final arithmetic intensity of our bootstrapping procedure is similar to what can be supported by state-of-art CPUs. 
Note that the addition of modular arithmetic vector extension to existing vector engines would already result in the overall application being memory bottlenecked.

\paragraph*{Datacenter GPUs} 
For GPU analysis we consider the NVIDIA A100 datacenter GPU. 
This GPU offers a peak $19.5$~TOp/s $32$-bit Integer FMA performance when clocked at $1.41$~GHz. 
It has an on-chip $40$~MB L2 last-level cache and uses an HBM2 DRAM interface supporting $1.55$~TB/s of bandwidth. 
Note again that a single die cannot fit a complete ciphertext in memory.
Applications with an INT32 FMA arithmetic intensity lower than $\sim 12$ will tend to be memory bottlenecked. 
In addition, $64$-bit integer arithmetic is not natively supported on a datacenter GPU and must be emulated in assembly which has a significant overhead (up to $20$ instruction for $64$-bit Integer multiply). 
As such for GPU implementations, it is advisable to use an RNS representation with $32$-bit limbs to avoid this overhead. 
Addition of native $32$-bit modular multiplication to future GPU will further worsen the memory bottleneck.

\emph{While the above estimations are simplistic and do not take into account the intricacies of instruction scheduling, the underlying point remains that raw access to compute power is not what bottlenecks existing FHE implementations. 
Building new hardware that merely adds an order-of-magnitude to the compute capability is unlikely to give an order of magnitude performance improvements without addressing the memory side of the story.}
\section{Evaluation}
\label{section:eval}
In this section, we compare our bootstrapping algorithm to prior art to demonstrate the improved throughput achieved by our optimizations. 
In addition, we show how our improved CKKS subroutines directly result in more efficient HE applications.

\subsection{Maximizing Bootstrapping Throughput}

\paragraph*{Bootstrapping Throughput}

Our metric to evaluate bootstrapping performance is based on the \emph{bootstrapping throughput} metric of  Han and Ki~\cite{HK19}. 
This metric attempts to capture the effectiveness of a bootstrapping routine by improving with the number of slots the algorithm bootstraps (which is the number of plaintext slots $n$), the number of limbs $\ell$ in the resulting ciphertext (which translates to the number of compute levels supported by the ciphertext), and the bit-precision $\bp$ of the plaintext data. 
These factors are then divided by the runtime of the bootstrapping procedure, denoted as $\brt$. This gives us the throughput metric in \Cref{eq:bootThroughput}.
\begin{equation}\label{eq:bootThroughput}
    \throughput = \frac{n \cdot \ell \cdot \bp}{\brt} 
\end{equation}

\paragraph*{Optimal Bootstrapping Parameters}
Given the throughput metric from \Cref{eq:bootThroughput}, we can select parameters to optimize it. 
We employ our architectural modeling tool to explore the parameter space of bootstrapping to maximize the throughput. 
As DRAM transfer times dominate in bootstrapping, our architectural model accounts for DRAM transfer time in the total runtime analysis, resulting in parameters that minimize DRAM transfers. 
The throughput-maximizing parameters for our fully-optimized bootstrapping algorithm (with all optimizations from \Cref{sec:cachingopts} and \Cref{sec:algoopts}) are given in \Cref{tab:bootParams}. 

\begin{table}[!ht]
    \centering
    \caption{Bootstrapping Parameters\\\emph{The $L$ parameter denotes the number of limbs in the ciphertext after the initial $\ModUp$ procedure in $\Bootstrap$. The $\fftIter$ parameter is the number of $\PtMatVecMult$ iterations in the $\CoeffToSlot$ and $\SlotToCoeff$ phases in $\Bootstrap$. The radix values for these iterations are all balanced, with any values that need to be larger placed at the end. The $\lambda$ value is the bit-security level.}}
    \label{tab:bootParams}
    \begin{threeparttable}
    \begin{tabular}{cccccccccc}
    \toprule
         && $L$ && $\dnum$ && $\fftIter$ && $\lambda$\\
    \midrule
        \textbf{Baseline} && $35$ && $3$ && $3$ && $<128$\tnote{$\dag$} \\
        \textbf{Best-case} && $40$ && $2$ && $6$ && $128$ \\
    \bottomrule
    \end{tabular}
    \begin{tablenotes}
    \item[$\dag$] The baseline set is based on ~\cite{GPUBoot21} originally targeting $\lambda = 128$. Updated cryptanalysis in ~\cite{BMTH20} reduces the security level for sparse keys. The parameters in this work include these updated recommendations. 
    \end{tablenotes}
    \end{threeparttable}
\end{table}

\subsection{Bootstrapping Performance Comparisons}
We now compare the throughput of our most optimized bootstrapping algorithm to prior art. 
To compare our algorithm to prior works, we re-implemented each algorithm in our architecture model. 
We then took the parameters given in each of these works and ran the algorithm in our model with these parameters. 
This allowed us to measure the total operations as well as the DRAM transfer times for each of these algorithms. 

From this analysis as well as our discussion in \Cref{sec:algoOptTakeaway}, we know that all of these bootstrapping algorithms are bottlenecked by the memory bandwidth. 
Therefore, we used the memory bandwidth requirement of each of these algorithms as a proxy for the overall runtimes. 
The memory requirement was converted to DRAM transfer time based on the memory bandwidth of the NVIDIA Tesla V100~\cite{NVIDIAV100}, which is $900$~GB/s.

The results of this analysis is presented in \Cref{tab:bootstrapping_comp}.
We now discuss each comparison in more detail. 
The parameter set selected from Jung et al.~\cite{GPUBoot21} is the same parameter set used as the baseline comparison in \Cref{sec:cachingopts}, which is the parameter set they give for their logistic regression implementation. 

We selected the parameter set from Bossuat et al.~\cite{BMTH20} that maximized their throughput. 
Note that this parameter set maximized the throughput when the runtime was measured on a CPU. 
For our architecture model, we are considering the case where computation has been accelerated to the point where runtime is completely dominated by memory transfers. 

The throughput computation for Samarzdic et al.~\cite{F1Paper21} was computed slightly differently since this work gives the DRAM bandwidth of their algorithm. 
However, this work only gives benchmarks for unpacked CKKS bootstrapping (i.e., there is no slot packing and the ciphertext only holds one element).
Rather than re-implementing their algorithm, we use the memory bandwidth usage they give for their unpacked CKKS bootstrapping, which is $721$~MB. 
To compute the runtime, we also use the peak DRAM bandwidth provided by the authors for their architecture, which is $1$~TB/s.
Using these two numbers, we found their bootstrapping procedure runtime to be $0.721$ milliseconds leading to the throughput number mentioned in \Cref{tab:bootstrapping_comp}.

\begin{table}[ht]
    \centering
    \caption{Bootstrapping comparison\\\emph{This table measures the bootstrapping throughput. The \textbf{Throughput} column is computed using \Cref{eq:bootThroughput} with the DRAM transfer time as a proxy for the runtime. The DRAM transfer time is measured in microseconds.}} 
    \begin{tabular}{cccccc} \toprule 
        \textbf{Work} & $n$ & $\ell$ & $\bp$ & \specialcell{\textbf{DRAM}\\\textbf{ Transfers}\\\textbf{(in GB)}} & \textbf{Throughput} \\
        \midrule 
        Jung et al.~\cite{GPUBoot21} & $2^{16}$ & $20$ & $19$ & $193.09$ & $116.07$ \\ \midrule
        Bossuat et al.~\cite{BMTH20} & $2^{15}$ & $16$ & $19$ & $75.30$ & $119.05$ \\  \midrule
        Samarzdic et al.~\cite{F1Paper21} & $1$ & $13$ & $24$ & $0.721$ & $0.43$ \\ \midrule
         Our Best Throughput & $2^{16}$ & $19$ & $19$ & $45.33$ & $469.68$ \\
        \bottomrule
    \end{tabular}
    \label{tab:bootstrapping_comp}
\end{table}

\subsection{Application Comparison}

A faster bootstrapping algorithm directly results in faster HE applications. Continuing with our running example of logistic regression training, we give benchmarks of the logistic regression algorithm from \Cref{section:FHEApplications} using our optimized bootstrapping routine and parameters. These benchmarks are given in \Cref{tab:building-block-improvement}.

\begin{table}[ht]
     \centering
     \caption{Performance of Logistic Regression Training Example\\\emph{This table displays benchmarks of the logistic regression bootstrapping application using our optimized bootstrapping parameters. In parentheses next to each benchmark, we give the improvement over \Cref{tab:building-block-cost}. }}
     \label{tab:building-block-improvement}
     \begin{tabular}{cccc}
     \toprule
     \specialcell{\textbf{Sub-routine}\\\textbf{Name}} & \specialcell{\textbf{Total}\\\textbf{Operations}\\\textbf{(in GOP)}} & \specialcell{\textbf{Total}\\\textbf{DRAM }\\\textbf{Transfers}\\\textbf{(in GB)}} & \specialcell{\textbf{Arithmetic}\\\textbf{Intensity}\\\textbf{(in Op/byte)}}\\
     \midrule
     $\InnerProduct$ &  $6.8256(1.2\times)$ & $3.3261(4.9\times)$ & $2.05(4.4\times)$ \\
     \midrule
     $\PolyEval$ & $2.2569(1.3\times)$ & $0.9745(3.6\times)$ & $1.97 (2.4\times)$ \\
     \midrule
     Full LR Iteration & $77.3846(1.2\times)$ & $41.0811(4.7\times)$ & $1.88(4\times)$ \\
     \midrule
     $\Bootstrap$ & $79.2401(1.9\times)$ & $45.3341(4.6\times)$ & $1.75(2.43\times)$ \\
     \bottomrule
     \end{tabular}
\end{table}

\subsection{Key Takeaways}
In this section, we demonstrated that our optimizations, which mostly focus on improving the arithmetic intensity of bootstrapping and other CKKS building blocks, result in a much higher memory throughput than prior art that mostly focused on optimizing the compute throughput.
This shows that focusing on compute throughput overlooks a crucial bottle-neck in CKKS applications: the memory bandwidth. 
To improve the overall performance of many important CKKS applications such as bootstrapping and encrypted logistic regression training, the memory bandwidth must be directly optimized.
\section{Discussion}
\label{section:Discussion}

Despite our algorithmic and cache optimizations to CKKS FHE bootstrapping (see Sections~\ref{sec:cachingopts} and \ref{sec:algoopts}), our analysis reveals that FHE bootstrapping continues to have low arithmetic intensity and is heavily bounded by main memory bandwidth.
This issue is not specific to CKKS bootstrapping alone. 
For one, bootstrapping algorithms for other FHE schemes such as BGV~\cite{BGV12} and B/FV~\cite{Brak12,FV12} have the same high-level structure and suffer from the same problem, although with different quantitative thresholds. 
Additionally, as discussed in Section~\ref{subsec:applConcCosts}, many natural applications (e.g. logistic regression and secure neural network evaluation) have the same high-level structure as bootstrapping, namely,  global linear operations followed by local non-linear operations, and consequently, they suffer from the main memory bottleneck as well.

Below, we discuss potential research avenues to solve this issue that is so central to the practicality of FHE.

\noindent \textbf{Future Improvements to Bootstrapping:} At a high-level, our optimizations can be viewed as improving the ``thrashing" of various low-level operations in the bootstrapping algorithm (as well as other natural applications of FHE such as encrypted training of machine learning models).
While future improvements may reduce thrashing in the baseline algorithms, the size of the ciphertexts and the size of the switching keys suggests that the overall arithmetic intensity is unlikely to drastically improve without a dramatic overhaul to FHE schemes.

In one extreme, we could be in the best-case-scenario for FHE bootstrapping.
In this world that we call ``FHE-mania'', all of bootstrapping can be done in cache without any DRAM reads or writes beyond the initial input and the final output. 
This world would call for true hardware acceleration of bootstrapping and would make our DRAM optimizations useless. 
On the other hand, we could be living in a world where the best possible bootstrapping algorithms remain bounded by the memory bandwidth.  
In this world that we call ``thrashy-land'', our optimizations remain crucial to achieving the highest throughput for bootstrapping. 
While it may be possible to optimize our way out of thrashy-land, as long as the RNS representation remains the dominant format of FHE data, our $\alpha$-limb and $\beta$-limb caching optimizations will remain relevant. 

A realistic possibility is a world that is somewhere in between FHE-mania and Thrashy-land. 
For example, it turns out that bootstrapping in GSW-like FHE schemes~\cite{GSW13,DM15} incurs slower noise growth and consequently smaller parameters $N$ and $Q$; however, it does not support packed bootstrapping as in BGV, B/FV and CKKS FHE schemes, a feature that is fundamentally important for efficiency. 
Can we achieve the best of both worlds? 
We believe there is exciting research to be done here (see \cite{MS18} for a  preliminary attempt); our analysis provides a compelling reason to pursue this line of research.

\noindent \textbf{Increase Main Memory Bandwidth:}
There are two approaches to increasing the main memory bandwidth.
First, we can use multiple DDRx channels, effectively using parallelism to increase the main memory bandwidth.
We could also use alternate main memory technologies like HBM2/HBM2e~\cite{Jun2017Hbm} that provide several times higher bandwidth than DDRx technology.
The second approach involves improving the physical interconnect between the compute cores and the memory by using silicon-photonic link technologies~\cite{Sun2015Nature}.
Judicious use of silicon-photonic technology can help improve the main memory bandwidth, and has the additional benefit of reducing the energy consumption for memory accesses.

\noindent \textbf{Improve Main Memory Bandwidth Utilization:} 
Here, there are two complementary approaches.
The first is to attempt a cleverer mapping of the data to physical memory to take advantage of spatial locality in cache lines such that we reduce the number of memory accesses required per compute operation. 
To complement this, we can improve FHE-based computing algorithms such that we perform more operations per byte of data that is fetched from main memory, i.e., improve temporal locality.
The second approach is algorithmic: namely, improve FHE bootstrapping algorithms (as discussed above) so that we reduce the size of the key-switching parameter, the main culprit for low arithmetic intensity, or eliminate it altogether.
These two complementary approaches may result in an increase in the arithmetic intensity, effectively reducing the time required for bootstrapping and FHE as a whole.

\noindent \textbf{Use In-Memory/Near-Memory Computing:}
Two potential architecture-level approaches include performing the operations in FHE APIs within main memory i.e., in-memory computing, and having a custom die very close to main memory for performing operations in FHE APIs, i.e., near-memory computing.
In the in-memory computing approach, we can eliminate a large number of expensive main memory accesses by performing matrix-vector multiplication operations in the main memory itself~\cite{Chi2016Isca}.
In contrast, in case of near-memory computing, we perform all the FHE compute operations in a custom accelerator that is placed close to the main memory.
Here, we cannot eliminate the memory accesses, but the cost of a memory access is lower than that of accessing a traditional memory.

\noindent \textbf{Use Wafer-Scale Systems:}
A radical technology-level solution is to design large-scale distributed accelerators such as Cerebras style wafer-scale accelerators~\cite{Cerebras} that have $40$~GB of high-performance on-wafer memory.
Tesla's Dojo accelerator~\cite{Tesla} also fits in this category wherein a large wafer is diced into $354$~chip nodes, which provides high bandwidth and compute performance.
Effectively, we can have large SRAM arrays i.e. large caches on the same wafer as the compute blocks, thus limiting all communication to on-chip wafer communication and avoiding expensive main memory accesses after the initial loads.
\section{Related Work}
\label{section:RelatedWork}

\noindent
\textbf{Algorithmic optimizations for CPUs:}
The key bottleneck in the FHE bootstrapping process is the large {\em homomorphic matrix-vector multiplication} required to convert ciphertexts from coefficient to evaluation representation and  back.
This requires many key-switching operations, which require accessing large number of switching keys from the DRAM, adding both to the computational cost and to data access latency. 
Initial implementations of bootstrapping in software (for example, the HEAAN library~\cite{CKKS17}) did try to reduce the number of rotations required in this linear transformation step by using baby-step giant-step (BSGS) algorithm, originally invented by Halevi and Shoup~\cite{HS18}. 
Using this algorithm, one can reduce the number of rotations to $O(\sqrt{N})$ while still requiring only $O(N)$ scalar multiplications.
The HEAAN library also optimizes the operational cost of approximating the modular reduction step by evaluating the sine function using a Taylor approximation. With these techniques, the HEAAN library takes about eight minutes to bootstrap $128$ slots within a ciphertext of degree $2^{16}$ on a CPU.

Chen, Chillotti and Song~\cite{CCS18} proposed a level collapsing technique along with BSGS for the linear transformation step to improve the number of rotations.
They also replaced the Taylor approximation with a more accurate Chebyshev approximation to evaluate a scaled-sine function instead. 
For the same parameter set as the HEAAN library, they observe a $3\times$ speedup. 
More recently, Han and Ki~\cite{HK19} proposed a hybrid key-switching approach to efficiently manage the amount of noise added through the key-switching operation. 
They evaluated a scaled, shifted cosine function instead of the scaled-sine function in modular reduction to reduce the number of non-scalar multiplications by half. 
Their optimizations led to an additional $3\times$ speedup. 
Bossuat et al.~\cite{BMTH20} further lower the operational complexity of the linear transformations by optimizing rotations through double-hoisting the hybrid key-switching approach. 
Double-hoisting the key-switch operation reduces the number of basis conversion operations significantly, which are expensive in terms of accessing the main memory. 
They also carefully manage the scale factors for non-linear transformations for error-less polynomial evaluation.
Their implementation in Lattigo library~\cite{lattigo} shows a further speedup of $1.5\times$ on a CPU.  

\noindent
\textbf{Algorithmic optimizations for GPUs:}
All the above mentioned optimizations heavily focused on lowering the operation complexity of bootstrapping, which led to a minor reduction in the main memory accesses as well. 
Recently, Jung et al.~\cite{GPUBoot21} presented the first ever GPU implementation of CKKS bootstrapping. 
Their analysis, even though limited to GPUs, rightly points out the main-memory-bounded nature of the bootstrapping operation. 
Thus, their optimizations, such as inter- and intra-kernel fusion, are all focused on improving the memory bandwidth utilization rather than accelerating the compute itself. 
Their bootstrapping implementation is so far the fastest requiring only $328.25$~ms (total time) for bootstrapping all the slots of a ciphertext of degree $N = 2^{16}$. 
As discussed in Section~\ref{sec:cachingopts} and \ref{sec:algoopts}, our techniques are composable with all these prior works and consequently, result in $3.2\times$ higher arithmetic intensity and $4.6\times$ reduction in main memory accesses.

\textbf{Hardware Accelerators for HE:} 
Samardzic et al.~\cite{F1Paper21} recently presented the architecture of a programmable hardware accelerator for FHE operations.
Their analysis also shows the fact that the FHE operations are memory bottlenecked. 
However, they implement a massively parallel compute block (having $4096$ modular multiplications) in their accelerator. 
From their performance analysis, it is evident that the compute block is underutilized due to the memory bottleneck. 


\section{Conclusion}
\label{section:Conclusion}
In this paper, we undertook a thorough architecture-level analysis of the compute and memory requirements for fully homomorphic encryption to identify the limits and opportunities for hardware acceleration.
Our analysis shows that the bootstrapping step is the critical performance bottleneck in FHE-based computing, and it has low arithmetic intensity and is heavily constrained by today's main memory systems.
We argue that to accelerate FHE-based computing, the research community should focus on improving the arithmetic intensity of FHE-based computing and leverage novel memory system architectures.
We proposed several architecture-independent and cache-friendly optimizations that improve arithmetic intensity by about $2.43\times$.
We also propose custom physical address mapping for \emph{limb-wise} and \emph{slot-wise} operations to enhance the main memory bandwidth utilization.
To further mitigate the impact of memory bandwidth on FHE-based computing, we suggest directions for the research community to explore novel techniques to either increase main memory bandwidth or improve its utilization, use in-memory/near-memory computing, and/or use wafer-scale systems with large on-chip memory.


\bibliographystyle{IEEEtranS}
\bibliography{bootstrapping-ckks-2022}

\appendices

\end{document}